\documentclass[letterpaper]{aa}  
\usepackage{graphicx}
\usepackage{natbib}
\usepackage{txfonts}
\newcommand{\degrees}[1]{\ensuremath{#1^\circ}}
\newcommand\Eav{{$\left<E\right>$}}
\newcommand\Fp{{$<\!\!F_p\!\!>$}}
\newcommand\Fi{{$<\!\!F_i\!\!>$}}

\begin{document}
   \title{The bright spiky emission of pulsar B0656+14}

   \author{P. Weltevrede
          \inst{1}
          \and
          G.~A.~E. Wright\inst{2,1}
	  \and
	  B.~W. Stappers\inst{3,1}
	  \and
	  J.~M. Rankin\inst{4,1}
          }

   \offprints{P. Weltevrede}

   \institute{Astronomical Institute ``Anton Pannekoek'',
              University of Amsterdam,
              Kruislaan 403, 1098 SJ Amsterdam, The Netherlands\\
              \email{wltvrede@science.uva.nl}
         \and
             Astronomy Centre, University of Sussex, Falmer, BN1 9QJ, UK\\
             \email{G.Wright@sussex.ac.uk}
	 \and
	     Stichting ASTRON, Postbus 2, 7990 AA Dwingeloo, The Netherlands\\
             \email{stappers@astron.nl}
	 \and
             Physics Department, 405 Cook Physical Science building, University of Vermont, Burlington, 05405, USA\\
             \email{Joanna.Rankin@uvm.edu}
             }

   \date{Received ...; accepted ...}

  \abstract
   { We present a detailed study of the single radio pulses of PSR
B0656+14, a pulsar also known to be a strong pulsed source of
high-energy emission. 
 }
   { Despite the extensive studies at high-energy wavelengths, there
is little or no published work on its single-pulse behaviour in the
radio band. In this report we rectify this omission. }
   { Radio observations using the Westerbork Synthesis Radio
Telescope at 1380 MHz and the Arecibo Observatory at 327 and 1525 MHz
are used to investigate the single-pulse behavior of PSR B0656+14. A
comparison is made with the phenomena of giant pulses and giant
micropulses. }
   { We have found that the shape of the pulse profile of
PSR B0656+14 requires\rm an unusually long timescale to achieve
stability (over 25,000 pulses at 327 MHz). This instability is caused
by very bright and narrow pulses with widths and luminosities
comparable to those observed for the RRATs. Many pulses are bright
enough to qualify as ``giant pulses'', but are broader than those
usually meant by this term.  At 327 MHz the brightest pulse was about
116 times brighter than the average pulse.  Although the most powerful
pulses peak near the centre of the profile, occasional sudden strong
pulses are also found on the extreme leading edge of the profile. One
of them has a peak flux of about 2000 times the average flux at that
pulse longitude. No ``break'' in the pulse-energy distributions is
observed, but nevertheless there is evidence of two separate
populations of pulses: bright pulses have a narrow ``spiky''
appearance consisting of short quasi-periodic bursts of emission with
microstructure, in contrast to the underlying weaker broad
pulses. Furthermore, the spiky pulses tend to appear in clusters which
arise and dissipate over about 10 periods. We demonstrate that the
spiky emission builds a narrow and peaked profile, whereas the weak
emission produces a broad hump, which is largely responsible for the
shoulders in the total emission profiles at both high and low
frequencies.  }
{}

   \keywords{Stars:pulsars:individual (PSR B0656+14) --- Stars:pulsars:general --- Radiation Mechanisms: non-thermal
               }

   \maketitle

\section{Introduction}

\object{PSR B0656+14}, at a spin-down age of 111,000 yrs, is one of three
nearby pulsars in the middle-age range in which pulsed high-energy
emission has been detected. These are commonly known as ``The Three
Musketeers'' (\citealt{bt97}), the other two being Geminga and PSR
B1055--52. Middle-aged pulsars (roughly defined to be those whose
spin-down ages range from 50,000 yrs to 300,000 yrs) are of interest
to neutron star theorists because they allow the detection of
high-energy thermal radiation from the star's surface, and hence can
provide tests for models of surface cooling and atmosphere
composition. In more energetic young pulsars (the Crab) the pulsed
non-thermal emission is dominant, and completely masks the thermal
emission at all wavelengths. In cool, older pulsars blackbody surface
temperatures are expected to have fallen below the detection
threshold, with the likely exception of their heated polar caps (the
polar cap of one such older pulsar, B0943+10, having recently been
detected in soft X-rays (\citealt{zsp05}).

However pulsars such as B0656+14 are also of interest because they
provide tests between competing emission models for the highest energy
components. Their hard X-ray emission can variously be interpreted as
a product of outergaps located in the outer magnetosphere
(\citealt{cz99}), or as originating closer to the surface magnetic
poles (\citealt{hm98}). Fundamental to these discussions are the
determination of the geometric location of the various high-energy
peaks, and their relation to the radio emission peak, which is
generally taken to define the polar cap of the pulsar's dipole axis.
In the case of PSR B0656+14, a major clue is that observations from
optical to hard X-ray all show one or both of the two peaks always at
the same two phase longitudes
(\citealt{dcm+05,kmmh03,pzs02,ssl+05,zps96}), while the radio peak
falls almost, but not quite, halfway between these peaks.

PSR B0656+14 was included in a recent extensive survey of subpulse
modulation in pulsars in the northern sky (\citealt{wes06}). In the
single pulses analysed for this purpose the unusual nature of this
pulsar's emission was very evident, especially the ``spiky'' nature of
the subpulses. But what was most striking was that the pulsar
occasionally exhibited exceptionally powerful and longitudinally
narrow subpulses reminiscent of ``giant'' pulses, hitherto reported
for only a handful of pulsars, including the Crab pulsar
(\citealt{sr68}) and mostly young or millisecond pulsars
(e.g. \citealt{spb+04}). More recently, giant pulses have been
discovered in two old pulsars (\citealt{ke04}). If their presence
could be confirmed in PSR B0656+14, this would demonstrate that the
phenomenon might also be found in pulsars of intermediate age and
hence at any stage of a pulsar's lifetime.
 
We therefore set out to explore the full nature of PSR B0656+14's
pulse behaviour in the radio band. PSR B0656+14 has a period (0.385
sec) not greatly below the average for all pulsars and at radio
frequencies exhibits an apparently unremarkable single-peak integrated
profile. Yet at any wavelength the integrated profile of a pulsar
conceals as much information as it yields, and this pulsar has proved
to be no exception. It is remarkable that, despite the extensive
studies at high-energy wavelengths, there is little or no published
work on its single-pulse behaviour in the radio band. In this report
we rectify this omission, and, in addition to confirming the presence
of extremely bright pulses across the profile, demonstrate that the
pulsar's radio emission is far from that typical of older better-known
pulsars.

While this work was being prepared the discovery of RRATs (Rotating
RAdio Transients) was announced. These are sources which emit single
powerful pulses, separated by long intervals, and were identified as
isolated pulsars with periods of between 0.4 and 7 seconds. However,
their relation to the known pulsar population was unclear.  The
intermittent giant pulses we have detected in PSR B0656+14 led us to
argue in a separate paper that this pulsar, were it not so near, could
itself have appeared as an RRAT (\citealt{wsr+06}). This paper is
complementary to that work, showing that this pulsar's bright pulses
are narrow and ``spiky'', occur at a wide range of central longitudes
and can be differentiated from a weaker but steadier underlying
emission. A third paper is planned in which we will present
polarization data and attempt to link the radio emission -- in a
single geometric structure -- to the high-energy peaks.

The details of the radio observations are described in the next
section. In Sect. \ref{SctProfile} the pulse profile is discussed, in
Sect. \ref{SctSinglePulse} the properties of the single pulses and in
Sect. \ref{SctSeparation5} we will show that the emission can be
decomposed into ``spiky'' and ``weak'' emission. In Sect.
\ref{SctDiscussion} our results are discussed and summarized.

\section{\label{SctObs}Radio observations}

\subsection{Introduction}

Initially, a relatively short sequence of about two thousand pulses of
PSR B0656+14 were collected with the Westerbork Synthesis Radio
Telescope (WSRT) to analyze its subpulse modulation properties as part
of a large survey for subpulse modulation at 21 cm
(\citealt{wes06}). It was found that the intensity of the single
pulses is modulated with frequencies larger than about 10 pulse
periods and the modulation index $m$ was found to be high, indicating
that the intensity variation of the single pulses is unusual.  The
single pulses of this pulsar would have not been further investigated
had not one of them triggered our interest: a single exceptionally
bright pulse on the leading edge far from the centre of the pulse
profile, which quickly led us to realise that the emission of the
pulsar was far from normal. This pulsar was then observed by the WSRT
for five more hours (of which two proved highly suitable because of
fortunate interstellar scintillation conditions). Archival Arecibo
Observatory (AO) data was also studied as well as a follow-up
observation. The details of the observations used in this paper can be
found in Table \ref{UsedObservationsTable} (including a reference key,
which we use throughout this paper).

\begin{table}
\caption{\label{UsedObservationsTable}The details of the observations
used in this paper. Here REF is the reference key used in this paper,
MJD is the modified Julian date of the observation, $\nu$ is the
central frequency of the observation, $\Delta\nu$ the bandwidth,
$\tau_\mathrm{samp}$ the sampling time and $N$ the number of recorded
pulses.}
\begin{center}
\begin{tabular}{r@{-}l|cr@{}lr@{}lr@{.}lc}
\hline
\hline
\multicolumn{2}{c|}{\hspace*{3.5mm}REF} & MJD & \multicolumn{2}{c}{$\nu$} & \multicolumn{2}{c}{$\Delta\nu$} & \multicolumn{2}{c}{$\tau_\mathrm{samp}$} & $N$ \\
\multicolumn{2}{c|}{} &  & \multicolumn{2}{c}{(MHz)} & \multicolumn{2}{c}{(MHz)} & \multicolumn{2}{c}{(ms)} &  \\
\hline
AO&P1  & 52840 & \hspace*{0.8mm}327&  & \hspace*{2mm}25&  & 0&5125 & 24765\\
AO&P2  & 53490 & \hspace*{0.8mm}327&  & \hspace*{2mm}25&  & 0&650  & 16888\\
AO&L   & 52854 & \hspace*{0.8mm}1525& & \hspace*{2mm}100& & 0&5125 & 15589\\
WSRT&L & 53437 & \hspace*{0.8mm}1380& & \hspace*{2mm}80&  & 0&2048 & 18586\\
\hline
\end{tabular}
\end{center}
\end{table}

\subsection{\label{SctClean}Cleaning of the data}

The WSRT data was of good quality without any Radio Frequency
Interference (RFI) to be removed, but the AO data did suffer from some
RFI. Pulses contaminated by RFI could relatively easily be identified
by calculating the root-mean-square (RMS) of the off-pulse noise. From
all AO observations the pulses with the highest off-pulse RMS (about
1\% of the data) were identified and excluded from further
analysis. The AO-P2 observation is, although shorter, of better
quality than the older AO-P1 observation.

The AO-L observation was affected by lightning, which caused spikes in
the data. To avoid confusing lightning with bright pulses from the
pulsar, we had to identify and remove the artificial spikes. This was
done by checking the data for the presence of non-dispersed spikes in
the frequency band. Using this method we were able to identify 46
spikes from the data which were removed by replacing them by samples
with the running baseline values, after which the data was
de-dispersed.

\section{\label{SctProfile}The shape and stability of the pulse profile}

\subsection{Shape of the pulse profiles}

\begin{figure}[tb]
\begin{center}
\resizebox{0.8\hsize}{!}{\includegraphics[angle=270,trim=0 0 0 0,clip=true]{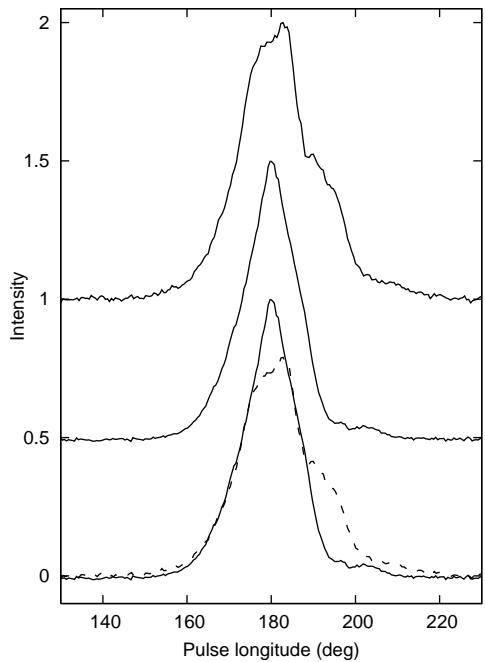}}\\
\end{center}
\caption{\label{ProfilesOverlayed}The top and middle profiles (solid
line) are the normalized pulse profiles of the 327-MHz AO-P1 and
1525-MHz AO-L observation respectively which are plotted with a
vertical offset of 0.5 and 1. The profiles are aligned such that the
central peak falls at pulse longitude \degrees{180} and the two
profiles are plotted overlaid at the bottom of the figure. The
intensity of the 327-MHz AO-P1 is scaled such that the central peaks
fit each other.}
\end{figure}

The average profiles of two of the AO observations are shown in
Fig. \ref{ProfilesOverlayed} and the top and middle profile are the
pulse profiles at two different frequencies. The shape of the pulse
profile of PSR B0656+14 is unusual for radio pulsars, especially at
high frequencies, where the profile is practically an isosceles
triangle. In fact, it proved impossible to fit this profile with any
Gaussian or Lorentzian form.

The profile at 1525 MHz is composed of an almost symmetric {\em
central peak} with a weak {\em shoulder} (from pulse longitude
$\sim$\degrees{190} to $\sim$\degrees{210}) on the trailing side of
the profile. At 327 MHz the pulse profile shape becomes more
complex. Although the profile is dominated by the central peak (the
leading part of the profile up to pulse longitude
$\sim$\degrees{190}), the shoulder (visible up to pulse longitude
$\sim$\degrees{200}) is more prominent and a {\em second shoulder}
appears (up to pulse longitude $\sim$\degrees{220}).  The central peak
seems to show some structure at 327 MHz (there is an excess at pulse
longitude $\sim$\degrees{182}), which is also visible in the AO-P2
observation. Note that the profiles are wide with a
full-width-half-maximum (FWHM) of \degrees{18.9} and \degrees{14.0} at
327 MHz and 1525 MHz respectively and a 10\% width to \degrees{39.4}
and \degrees{28.8} at 327 MHz and 1525 MHz respectively.

\subsection{\label{SctAlignment}Alignment of the pulse profiles}

It is important to keep in mind that the observations are not time
aligned, so the profiles of each observation have an
arbitrary pulse longitude offset. At the bottom of
Fig. \ref{ProfilesOverlayed} we have overlaid a 327 and 1525-MHz
profile, keeping the centroid of the central peak at a fixed pulse
longitude. The pulse profiles are scaled in intensity such that the
central component overlaps. Although the relative shift and scaling
are subjective, the fact that the profiles fit so accurately is
surprising and shows that the central component of the two profiles
maintains an invariant width over frequencies more than a factor four
apart.

This alignment suggests that the profile almost exclusively evolves
with frequency on the trailing side. The pulse profile becomes broader
at lower frequencies as expected from radius-to-frequency mapping
(RFM; \citealt{cor78}). However it is doubtful whether this is
actually the origin of the broadening. One could argue from the bottom
profiles of Fig. \ref{ProfilesOverlayed} that there is no RFM at all
in the central peak and that the broadening of the pulse profile is
only due to the appearance of more emission components.

\subsection{\label{SctStability}Stability of the pulse profile}

For most pulsars one can obtain a stable pulse profile by averaging a
few to a few hundred pulses, and the profiles of younger pulsars tend
to stabilize more quickly (\citealt{hmt75,rr95}), so our observations
of up to 25,000 pulses could have been expected to be long
enough. However PSR B0656+14 proved to be far from typical, and the
stability of its profile turns out to be an important issue in
determining the ``true'' shape of its pulse profile.

\begin{figure}
\resizebox{!}{0.76\hsize}{\includegraphics[angle=0,trim=0 0 0 15,clip=true]{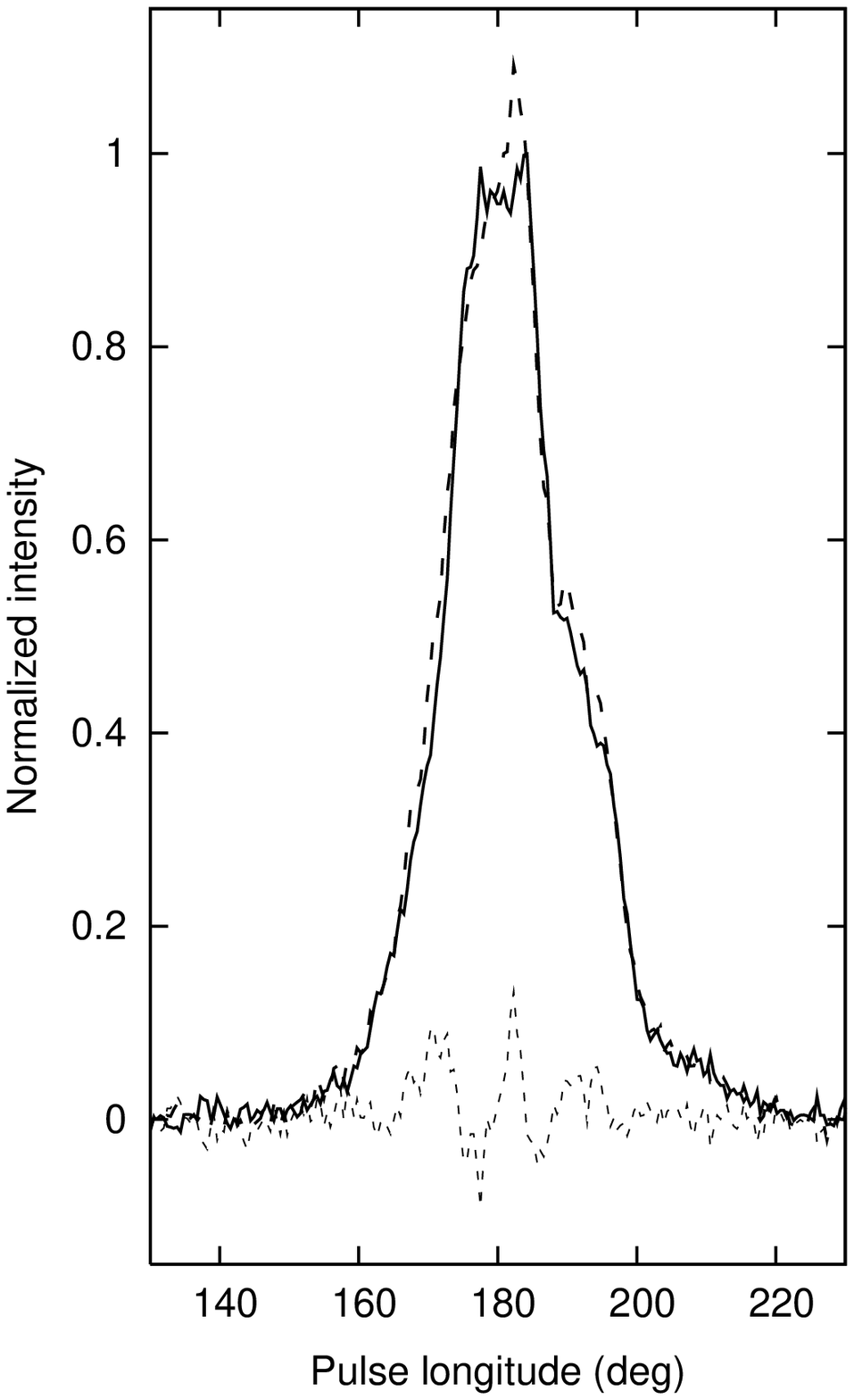}}
\resizebox{!}{0.76\hsize}{\includegraphics[angle=0,trim=0 0 0 0,clip=true]{5572f2b.ps}}\\
\centerline{}
\resizebox{!}{0.76\hsize}{\includegraphics[angle=0,trim=0 0 0 15,clip=true]{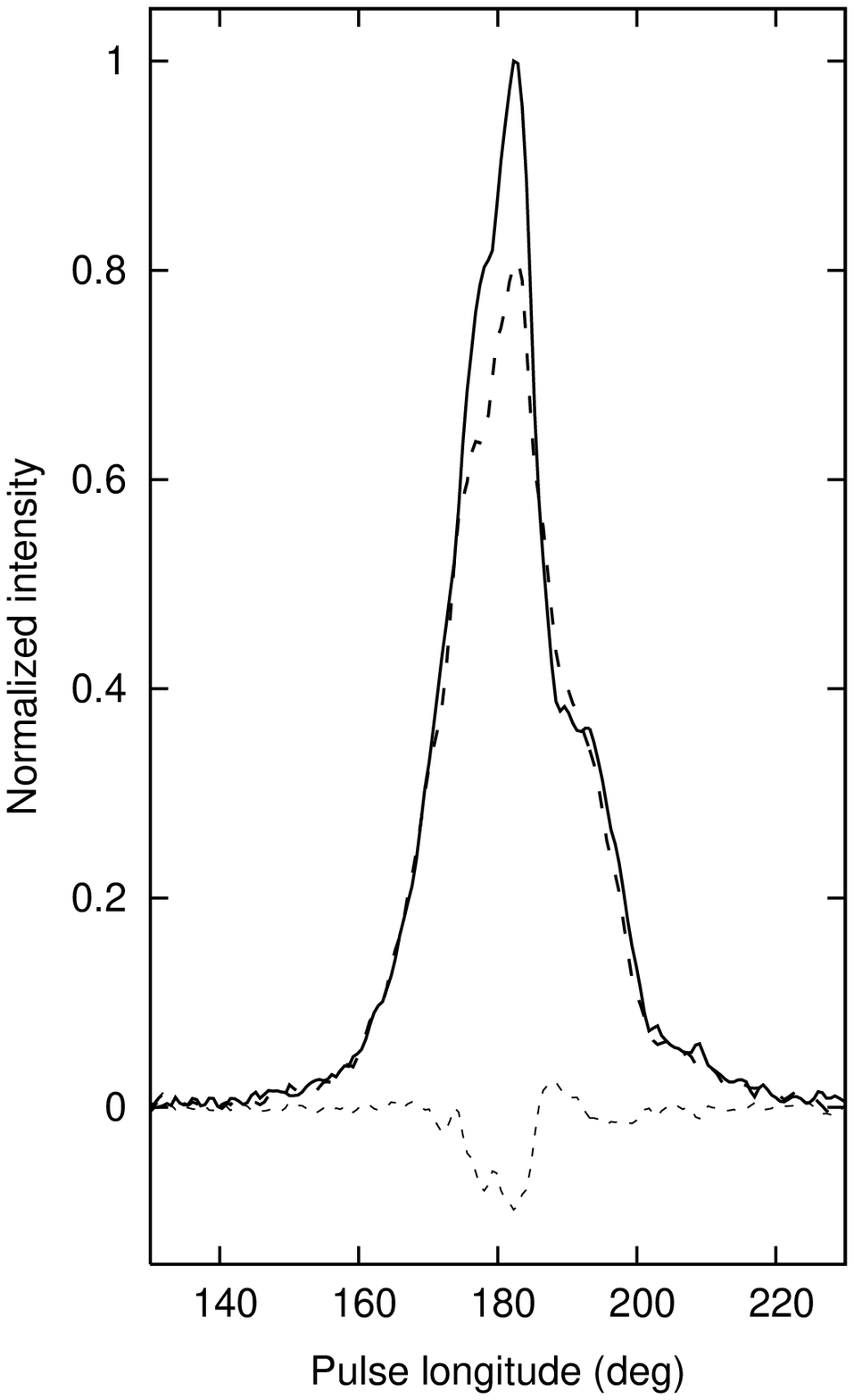}}
\resizebox{!}{0.76\hsize}{\includegraphics[angle=0,trim=0 0 0 0,clip=true]{5572f2d.ps}}\\
\centerline{}
\resizebox{!}{0.76\hsize}{\includegraphics[angle=0,trim=0 0 0 15,clip=true]{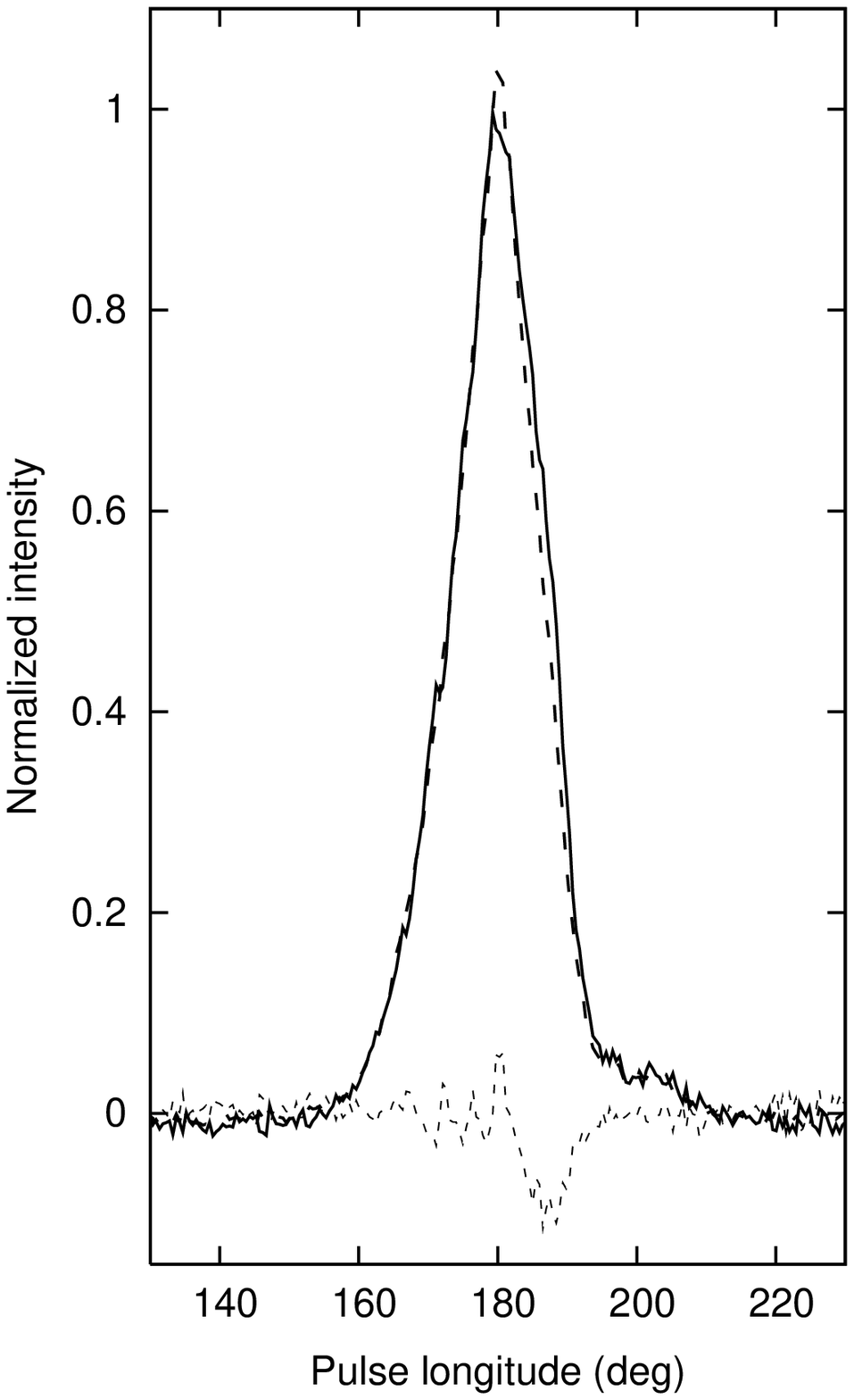}}
\resizebox{!}{0.76\hsize}{\includegraphics[angle=0,trim=0 0 0 0,clip=true]{5572f2f.ps}}
\caption{\label{Profiles}From top to bottom these panels show the
327-MHz AO-P1, AO-P2 and 1525-MHz AO-L observations. The overlaid
profiles of the left panels are the average-pulse profiles of the
first (solid line) and second halves (dashed line) of the observations
and the dotted lines are their differences. Only the two overlaid
profiles in the bottom left panel are scaled in intensity to fit each
other. The profiles of the right panels are the pulse profiles
obtained by averaging successive blocks of one thousand pulses each. }
\end{figure}

To test whether our data sets are long enough to get a stable pulse
profile, we split the data into two halves and computed their average
profiles. The overlaid profiles in the left panels of
Fig. \ref{Profiles} show the average profiles of the first and second
halves of two observations. As one can see, the profiles of the two
halves of the observation are similar, but with significant
differences. This is especially the case on the trailing side of the
1525-MHz pulse profile and in the central peak of the 327-MHz
profiles.

To further illustrate the time dependence of the pulse profile, the
profiles of successive blocks of one thousand pulses were calculated
(right panels of Fig. \ref{Profiles}). Unlike the 327-MHz profiles
(top and middle right panels), the 1525-MHz profiles (bottom right
panel) have different intensities because of interstellar
scintillation. This suggests that at 327 MHz the scintillation bandwidth
is much smaller than the observing bandwidth. In Sect. \ref{scintct}
we will argue that this is indeed the case. Note that apart from the
intensity changes due to scintillation the shape of the profile is
much more stable at high frequencies compared with low frequencies.

As one can also see in Fig. \ref{Profiles}, the profile shape changes
are very significant on timescales of one thousand pulses, especially
at 327 MHz. A much longer observation would be required to find out if
there exists a time scale for the pulse profile to stabilize. The
profiles in the top left panel of Fig. \ref{Profiles} are not scaled
in intensity to match and are unaffected by scintillation. This means
that the radio power output of the pulsar at 327 MHz is very stable,
although the profile shape is not. 

In the middle left panel of Fig. \ref{Profiles}, the pulse profiles of
the first and second halves of the AO-P2 observation are shown
overlaid. Unlike the AO-P1 observation (top left Fig. \ref{Profiles}),
the intensity has significantly changed during the observation. Note
that the intensity of the central peak has changed in height, while
the intensities of the two shoulders remains the same. The middle
right panel of Fig. \ref{Profiles} confirms that the pulse profile is
changing gradually around pulse 11,000 (block 11) from a more peaked
to a much broader shape. So not only is the profile highly unstable on
short timescales, it can also evolve gradually on a timescale of
hours.

It must be noted that a non-equatorial mounted telescope with
cross-coupling errors can introduce a gradual time-evolution in the
shape of (highly linear polarized) pulse profiles as a function of
parallactic angle.  However, the observations are calibrated for
polarization and the parallactic-angle rotation is taken into
effect\footnote{PSR B0656+14 if far enough away from the Arecibo
zenith to fully accommodate parallactic-angle rotation.}. Moreover,
the particular type of cross-coupling which occurs in the AO
instrument is well studied and has the property of preserving both
total and polarized power. Therefore, systematic errors due to
polarization calibration uncertainties are probably not responsible
for the observed effects.

\subsection{\label{scintct}Scintillation properties}

\begin{figure}[tb]
\resizebox{0.99\hsize}{!}{\includegraphics[angle=0,trim=0 0 0 0,clip=true]{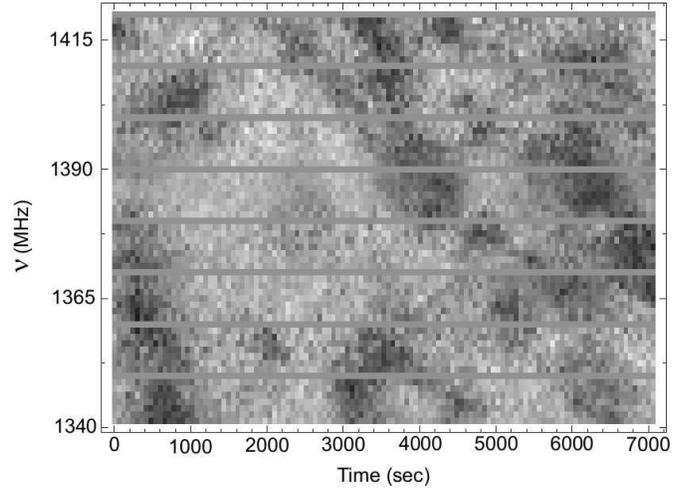}}
\caption{\label{DynamicSpectrum}The dynamic spectrum of the 1380-MHz
WSRT-L observation. Here the measured signal to noise of the pulsar
signal is plotted in grayscale as a function of both time and
frequency. The horizontal stripes in the dynamic spectrum
show where the eight frequency-bands roll off. } 
\end{figure}

The dynamic spectrum of the 1380-MHz WSRT-L observation is shown in
Fig. \ref{DynamicSpectrum}. One can see that the scintles (the dark
patches) have a characteristic timescale (the scintillation timescale)
of the order of a thousand seconds. The horizontal stripes in the
dynamic spectrum are because the 80-MHz bandwidth of the observation
is divided into eight 10-MHz bands. At the edges of the bands there is
no useful data because the bandpass rolls off just inside the
full-width of 10 MHz.  The scintillation bandwidth (the characteristic
frequency scale of the scintles) is clearly less than the total
bandwidth of the observation (80 MHz). In the thin screen
approximation, the scintillation bandwidth is proportional to $\nu^4$
(\citealt{sch68}), close to the observed scaling
(e.g. \citealt{bcc+04}). Using this scaling one expects a
scintillation bandwidth smaller than 0.3 MHz at 327 MHz. This is much
smaller than the observing bandwidth of the AO observations at that
frequency (25 MHz). Therefore the effect of the individual scintles
will be averaged out at 327 MHz, causing intensity fluctuations due to
scintillation to be invisible in the top and middle panels of
Fig. \ref{Profiles}. In fact, the scintillation bandwidth at this
frequency is so small that it is comparable, or even smaller than, the
the bandwidth of the individual frequency channels (0.098 MHz). This
means that it is not possible to measure it directly from our
observations, which is confirmed in the middle panel of
Fig. \ref{megapulse_disp}.

\section{\label{SctSinglePulse}The radio bursts of PSR B0656+14}

\begin{figure}[tb]
\begin{center}
\resizebox{0.75\hsize}{!}{\includegraphics[angle=0]{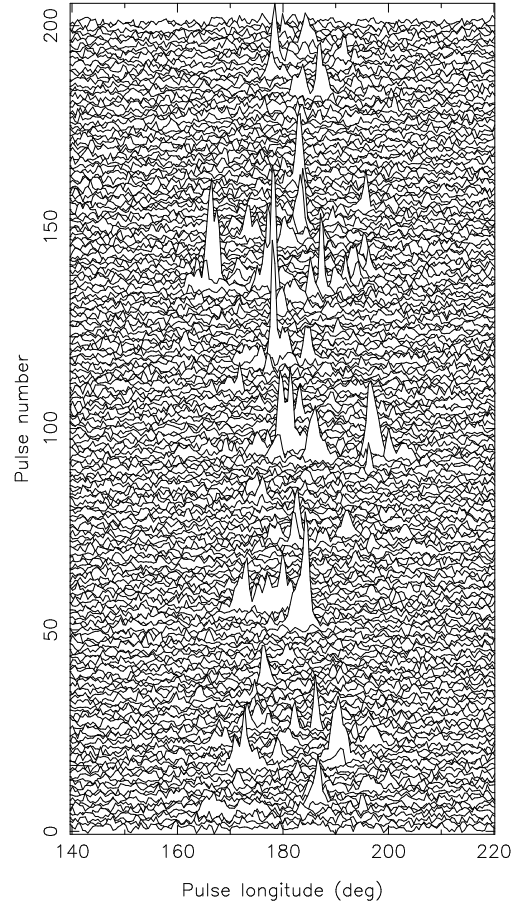}}
\end{center}
\caption{\label{Stack}A typical sequence of successive pulses of the
327-MHz AO-P1 observation.}
\end{figure}

A typical pulse sequence of this pulsar at 327 MHz is shown in
Fig. \ref{Stack}. The plotted pulse-longitude range is the whole range
within which the pulsar is found to emit (compare with
Fig. \ref{Profiles}). One can see that the frequent outbursts of radio
emission are much narrower than the width of the pulse profile. The
emission also has burst-like behaviour in the sense that the radio
outbursts tend to cluster in groups of a few pulse
periods. Furthermore, this clustering sometimes seems to be weakly
modulated with a quasi-periodicity of about 20 pulse periods (see for
instance the bursts around pulse numbers 55, 75, 95, 115 and
135). Apart from these bursts there are many pulses (and large
fractions of the pulse window) that contain no signal above the noise
level.

We will use the term {\em spiky} to refer to these bursts of radio
emission. In this section we investigate the energy distributions of
this pulsar, leading to comparisons with the phenomena of giant pulses
and giant micropulses.

\subsection{Correction for scintillation}

To compare the bright pulses of PSR B0656+14 with the giant pulse
phenomenon, we have calculated the pulse-energy distributions. Because
the dispersion measure (DM) is quite low for this pulsar,
scintillation can be severe. From Figs. \ref{Profiles} and
\ref{DynamicSpectrum} it is clear that we have to be careful in
interpreting the energy distributions of the high frequency
observations. Without correction, the pulse energies will be over- and
underestimated when interstellar scintillation makes the pulsar signal
respectively stronger and weaker. To correct for this broadening of
the pulse-energy distribution, the pulse energies of the two high
frequency observations are compared with the running average of the
pulse energies (which we will denote by {\Eav}), instead of simply
with the average-pulse energy of the whole observation.

We used a running average of 600 pulses. This timescale is chosen such
that the running average smoothly follows the scintillation (which has
a timescale of the order of one thousand seconds). This timescale is
also sufficiently large compared with the timescale of the
pulse-intensity modulation ($\sim20$ pulse periods, as we will show in
Sect. \ref{SctModulation}). To prevent the running average from being
influenced too much by very bright pulses, we do not include pulses
above 6 {\Eav} (the effect of this correction turns out to be
negligible).  The correction for scintillation will broaden the
off-pulse energy distribution. This is because when scintillation
makes the pulsar weaker, the noise RMS (expressed in terms of the
running-average pulse energy {\Eav}) becomes larger.  To improve the
signal-to-noise ($S/N$) ratio, a threshold for the running average is
used. If {\Eav} drops below this threshold, the pulses are not used
for the scintillation-corrected pulse-energy distribution.

\subsection{Giant pulses?}

\begin{figure*}[htb]
\begin{center}
\resizebox{0.32\hsize}{!}{\includegraphics[angle=0,trim=10 0 23 0,clip=true]{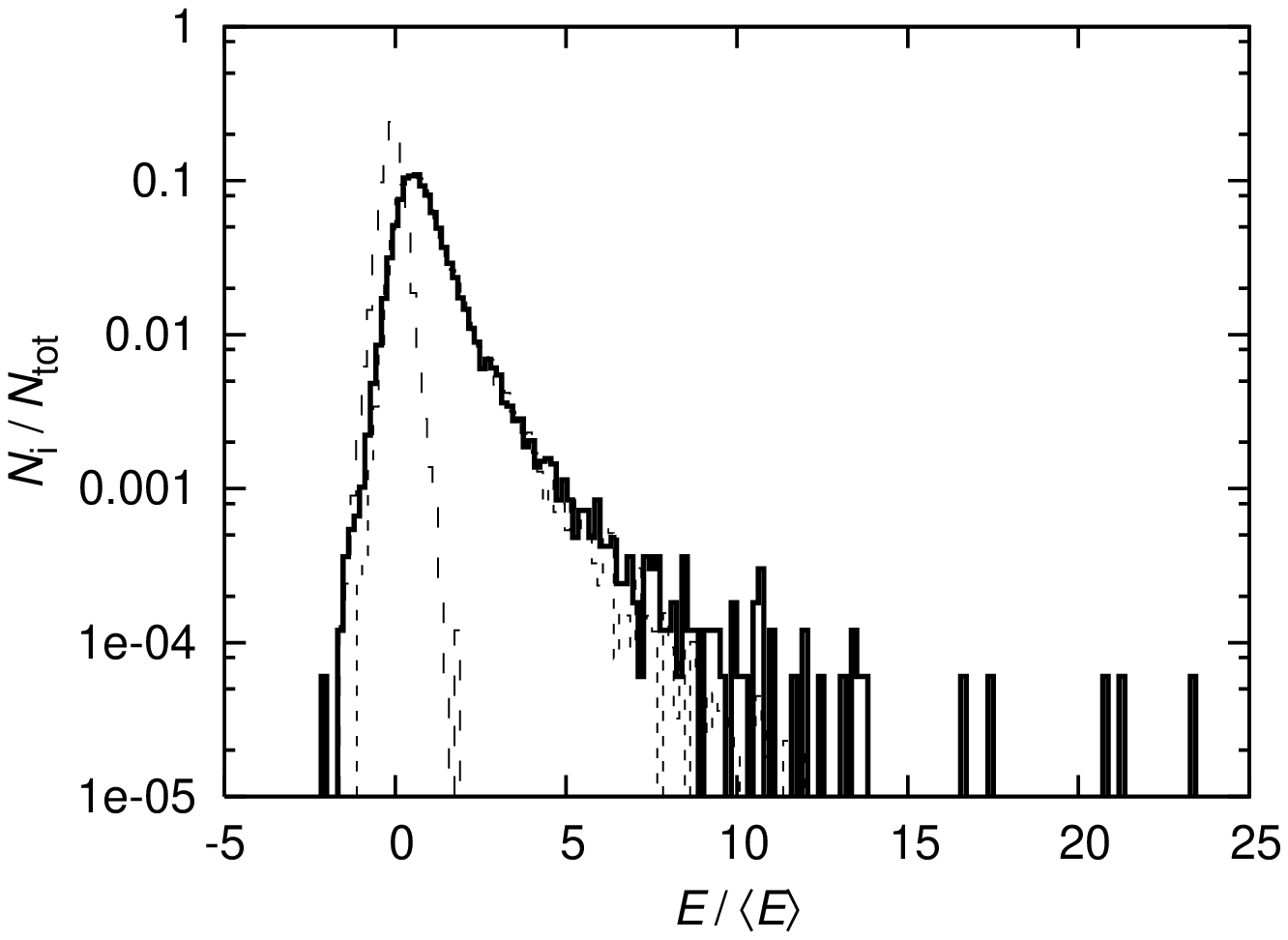}}
\hspace{0.01\hsize}
\resizebox{0.32\hsize}{!}{\includegraphics[angle=0,trim=10 0 23 0,clip=true]{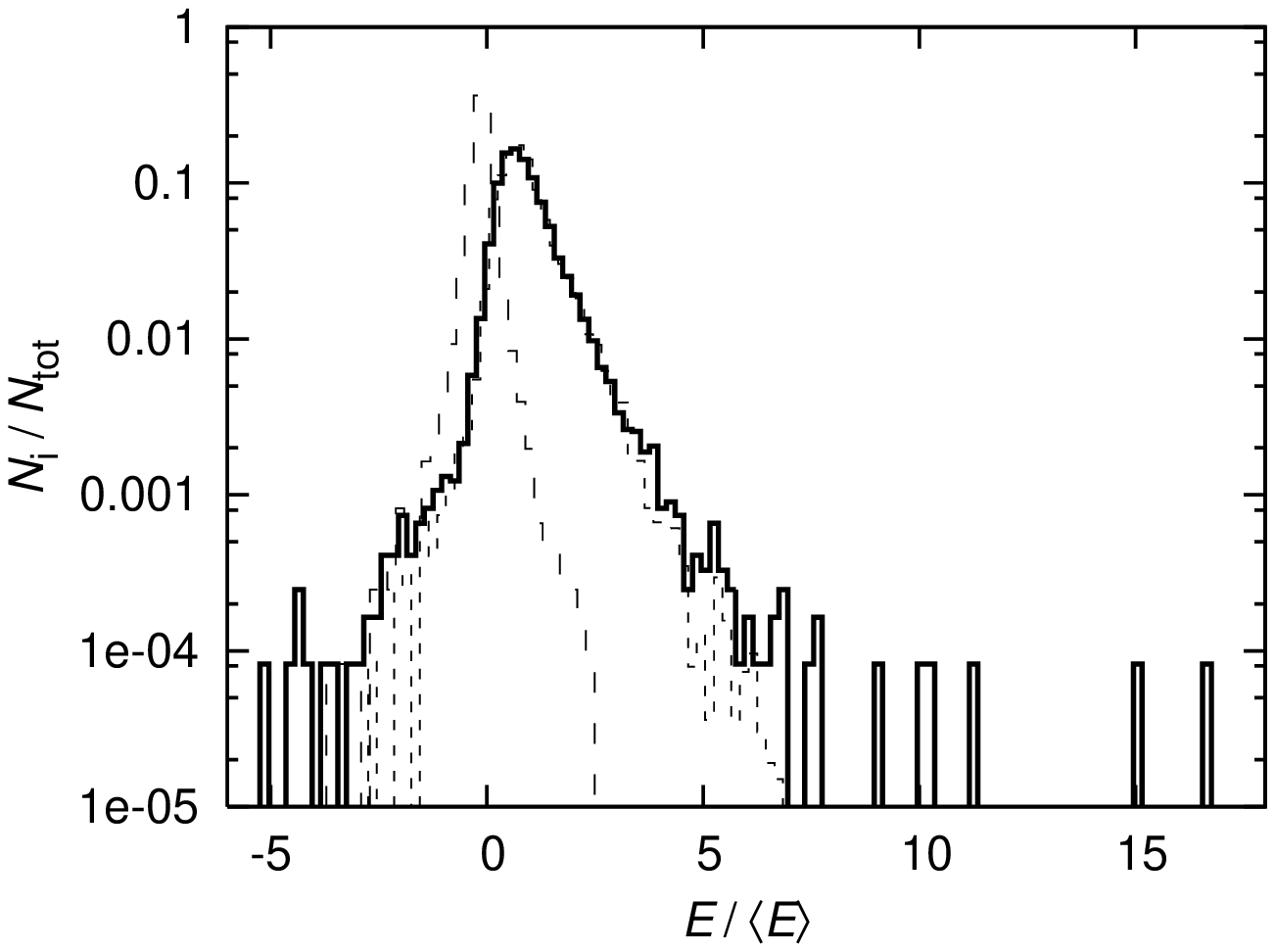}}
\hspace{0.01\hsize}
\resizebox{0.32\hsize}{!}{\includegraphics[angle=0,trim=10 0 23 0,clip=true]{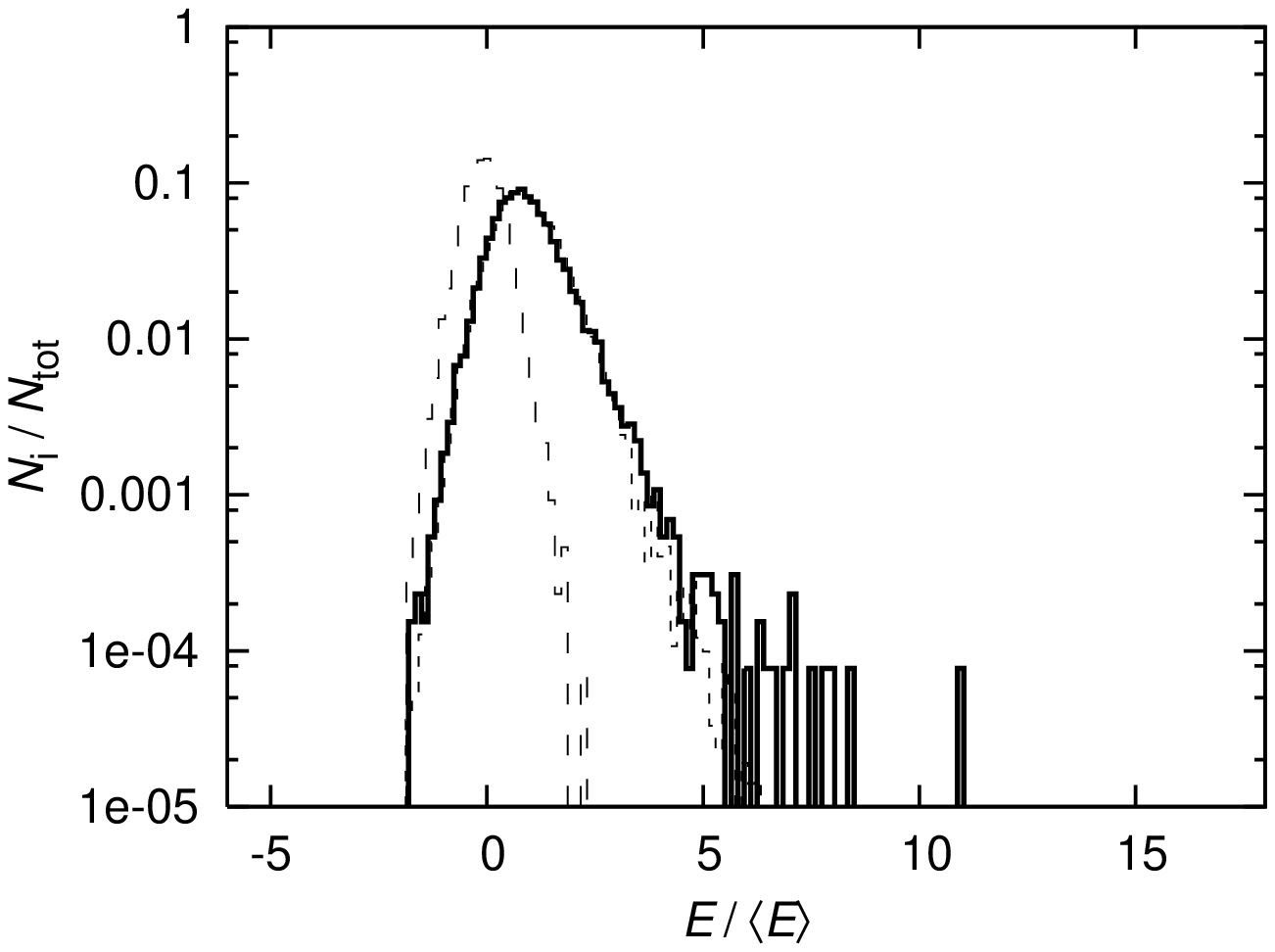}}
\end{center}
\caption{\label{Enhists}The pulse-energy distribution (solid line) of
(from left to right) the 327-MHz AO-P2, 1525-MHz AO-L and the 1380-MHz
WSRT-L observations, together with the fit (dotted line). The
off-pulse energy distributions are the dashed lines. The energies are
normalized to the average-pulse energy and the two high frequency
observations are corrected for scintillation. In the AO-P2 observation
there is a pulse of 116 {\Eav}, which falls outside the plotted
range.}
\end{figure*}

The scintillation-corrected high-frequency pulse-energy distributions
are shown in Fig. \ref{Enhists}, together with that for the 327-MHz
AO-P2 observation which did not require correction. The pulse-energy
distributions extend over an extremely wide energy range.  In the
AO-P2 observation the brightest pulse is about 116 {\Eav} (outside the
plotted energy range). This is well above the ``working definition of
giant pulses'' threshold of 10 {\Eav} (e.g. \citealt{cai04}).  In the
327-MHz AO-P1 and AO-P2 observation respectively 0.4\% and 0.2\% of
the pulses are above this threshold. In the 1525-MHz AO-L and in the
1380-MHz WSRT observation this percentage is an order of magnitude
less (0.04\% and 0.01\%).

\begin{figure}[tb]
\begin{center}
\resizebox{0.9\hsize}{!}{\includegraphics[angle=0]{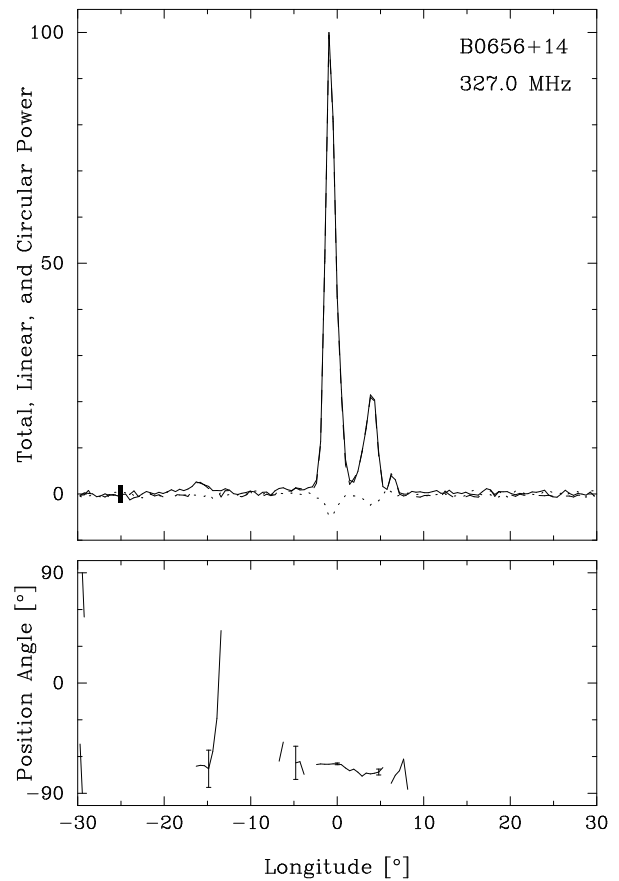}}
\end{center}
\caption{\label{polarization_plot} The top panel shows a typical
bright single pulse of the 327-MHz AO-P1 observation in total
intensity (solid line), linear polarization (dashed line, but almost
indistinguishable from the solid line) and circular polarization
(dotted line). The peak-flux of this pulse is normalized to
100. The bottom panel shows the polarization angle.}
\end{figure}

PSR B0656+14 therefore emits giant pulses in the sense that it emits
single pulses with energies above 10 {\Eav}. Another characteristic of
giant pulses is that they are very narrow with timescales down to
nano-seconds (e.g. \citealt{spb+04,hkwe03}). Also the giant pulses of
most pulsars are emitted in a small pulse-longitude range compared
with the pulse profile. In Fig. \ref{polarization_plot} one can see an
example of a bright pulse from PSR B0656+14 which shows they have
clear structure and are much broader than the ``classical'' giant
pulses. In order to determine whether the bright pulses are associated
with certain locations in the pulse profile, we have added the single
pulses within certain pulse-energy ranges (left panels of
Fig. \ref{lrced}). It is clear that the brightest pulses with $E>10$
{\Eav} are not strictly confined to a small pulse-longitude
range. Nonetheless they are limited to the leading and central
regions, i.e. the shoulders are clearly associated with the weak
pulses. Especially at 327 MHz there is a remarkably sharp boundary
between the central peak and the shoulder for the occurrence of the
bright pulses. The profile of the summed brightest pulses of the
327-MHz AO-P1 observation appears to have a separate component on the
leading side of the pulse profile (at pulse longitude
\degrees{165}). However only 15 pulses contribute to this component,
which is furthermore not visible in the AO-P2 observation, and is
therefore possibly not significant.

\subsection{Giant micropulses?}

\begin{figure*}[tb]
\begin{center}
\resizebox{!}{0.39\hsize}{\includegraphics[angle=0,trim=0 0 0 0,clip=true]{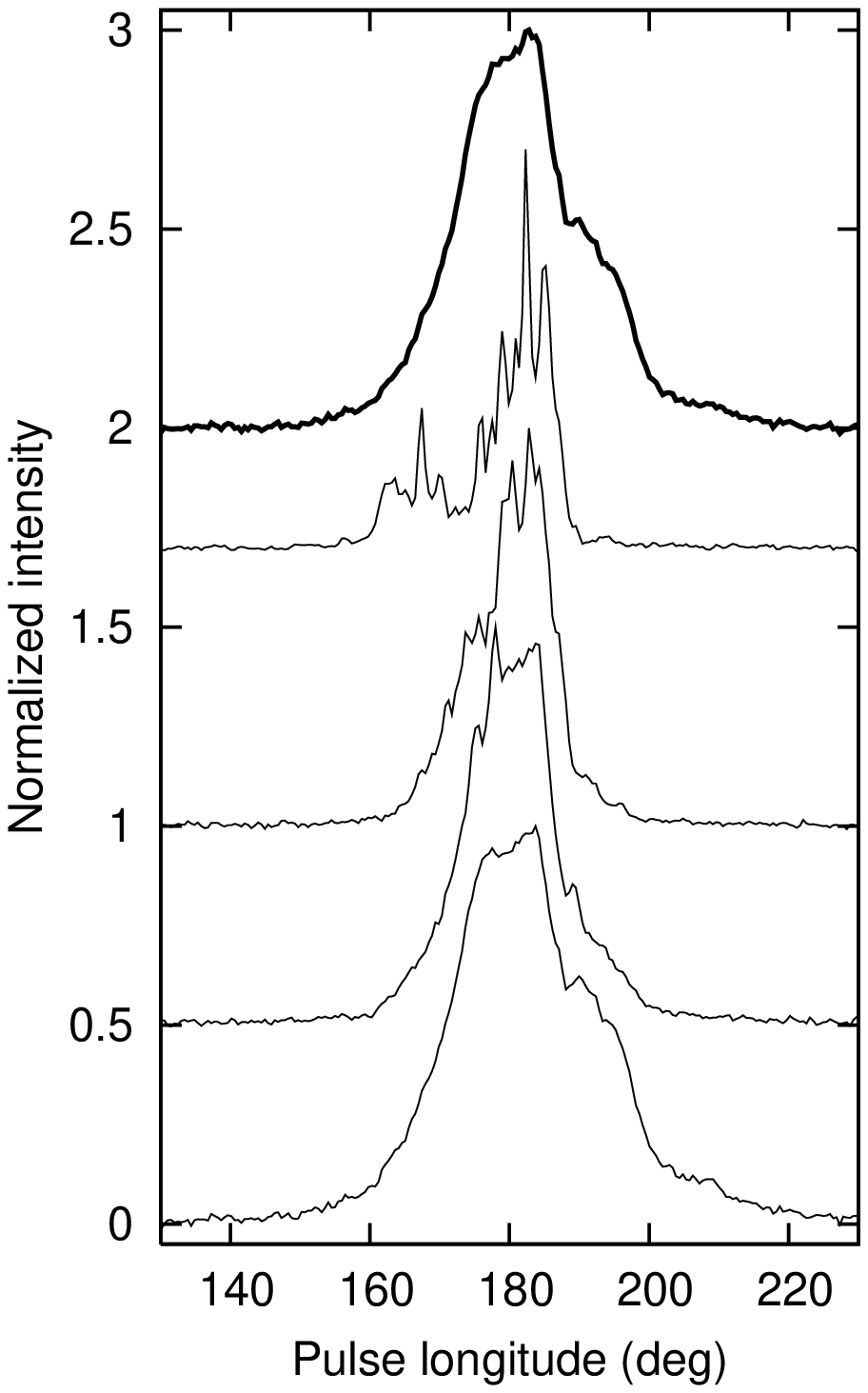}}
\resizebox{!}{0.39\hsize}{\includegraphics[angle=0,trim=0 0 0 0,clip=true]{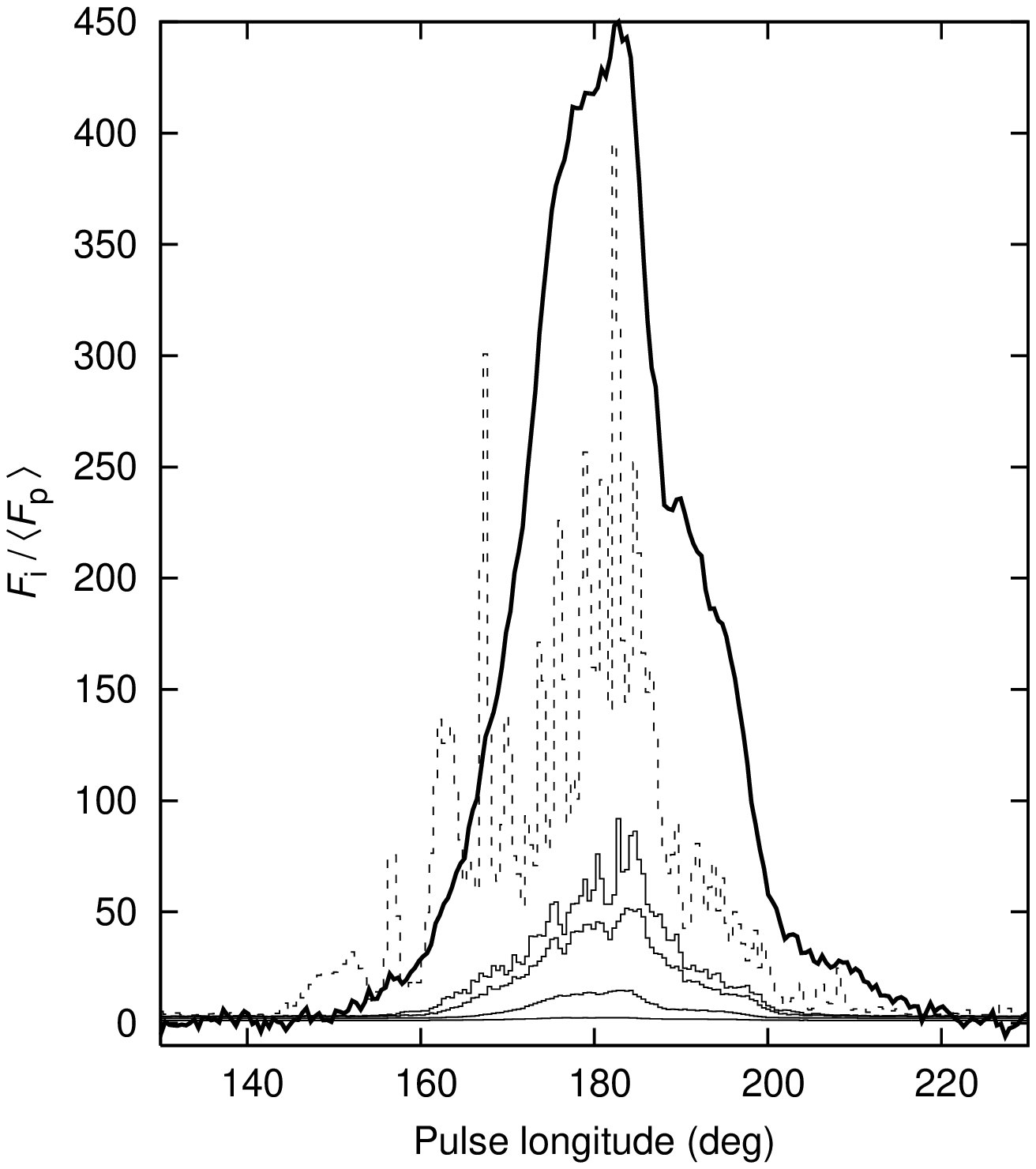}}
\resizebox{!}{0.39\hsize}{\includegraphics[angle=0,trim=0 0 0 0,clip=true]{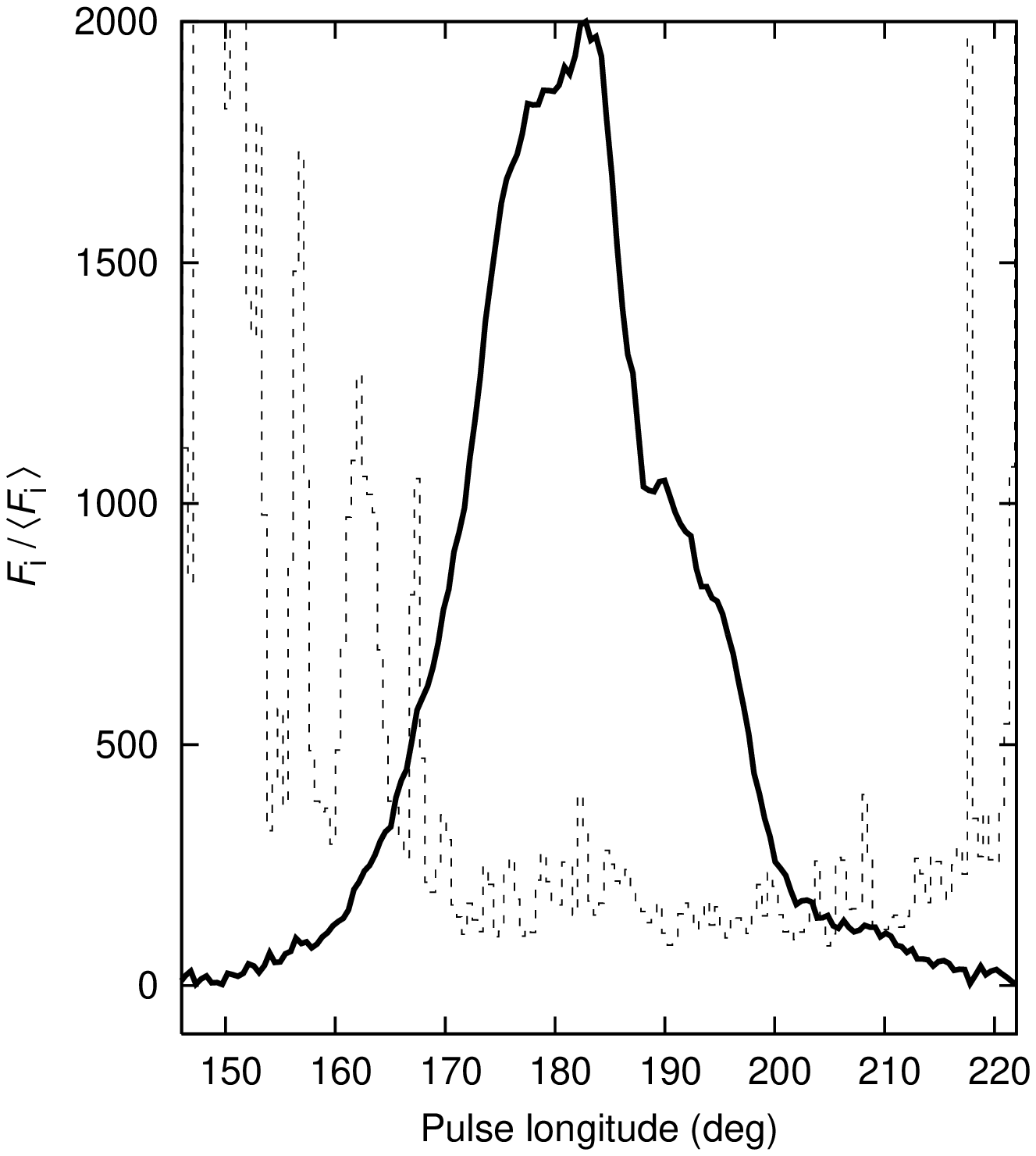}}\\
\resizebox{!}{0.39\hsize}{\includegraphics[angle=0,trim=0 0 0 0,clip=true]{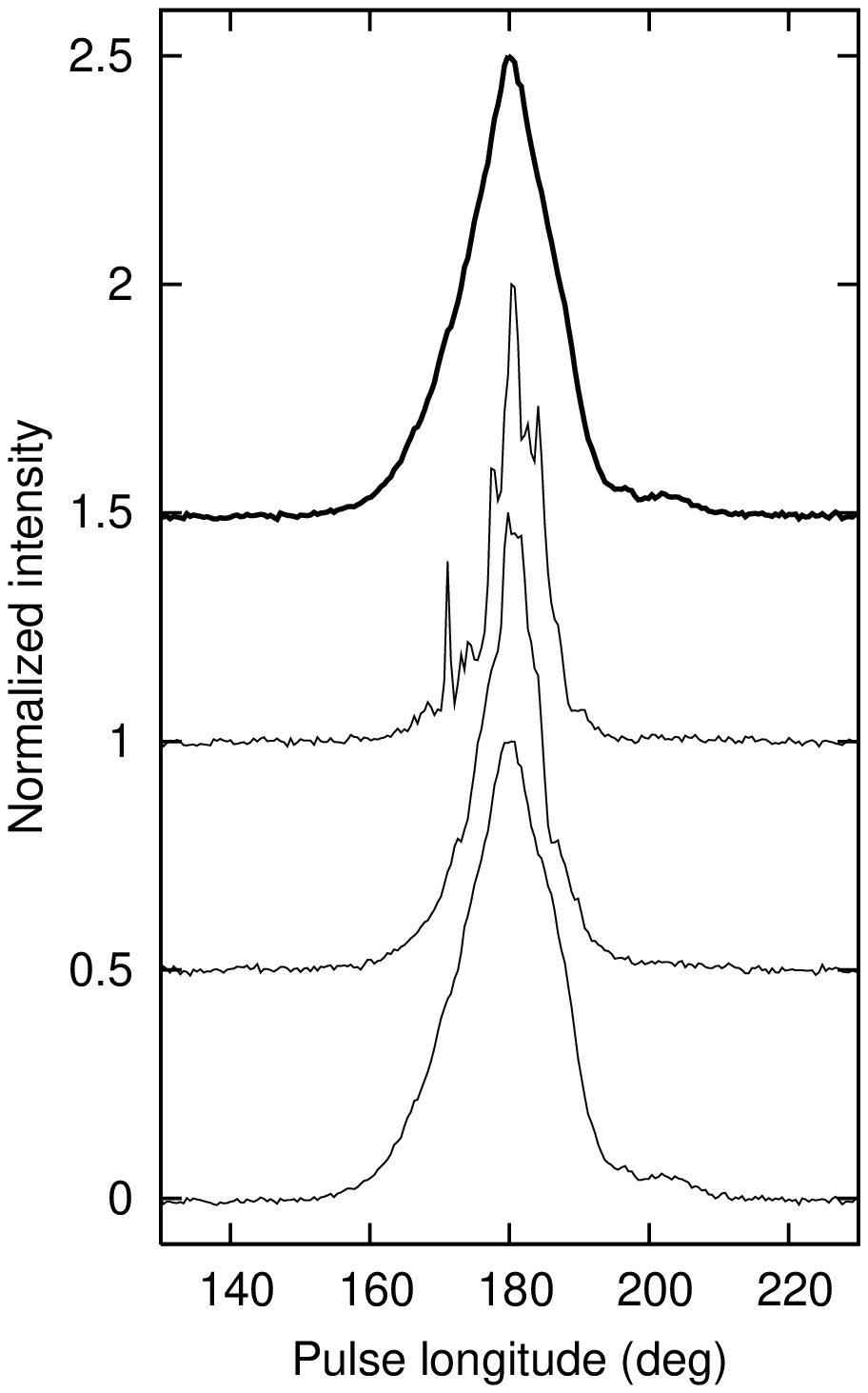}}
\resizebox{!}{0.39\hsize}{\includegraphics[angle=0,trim=0 0 0 0,clip=true]{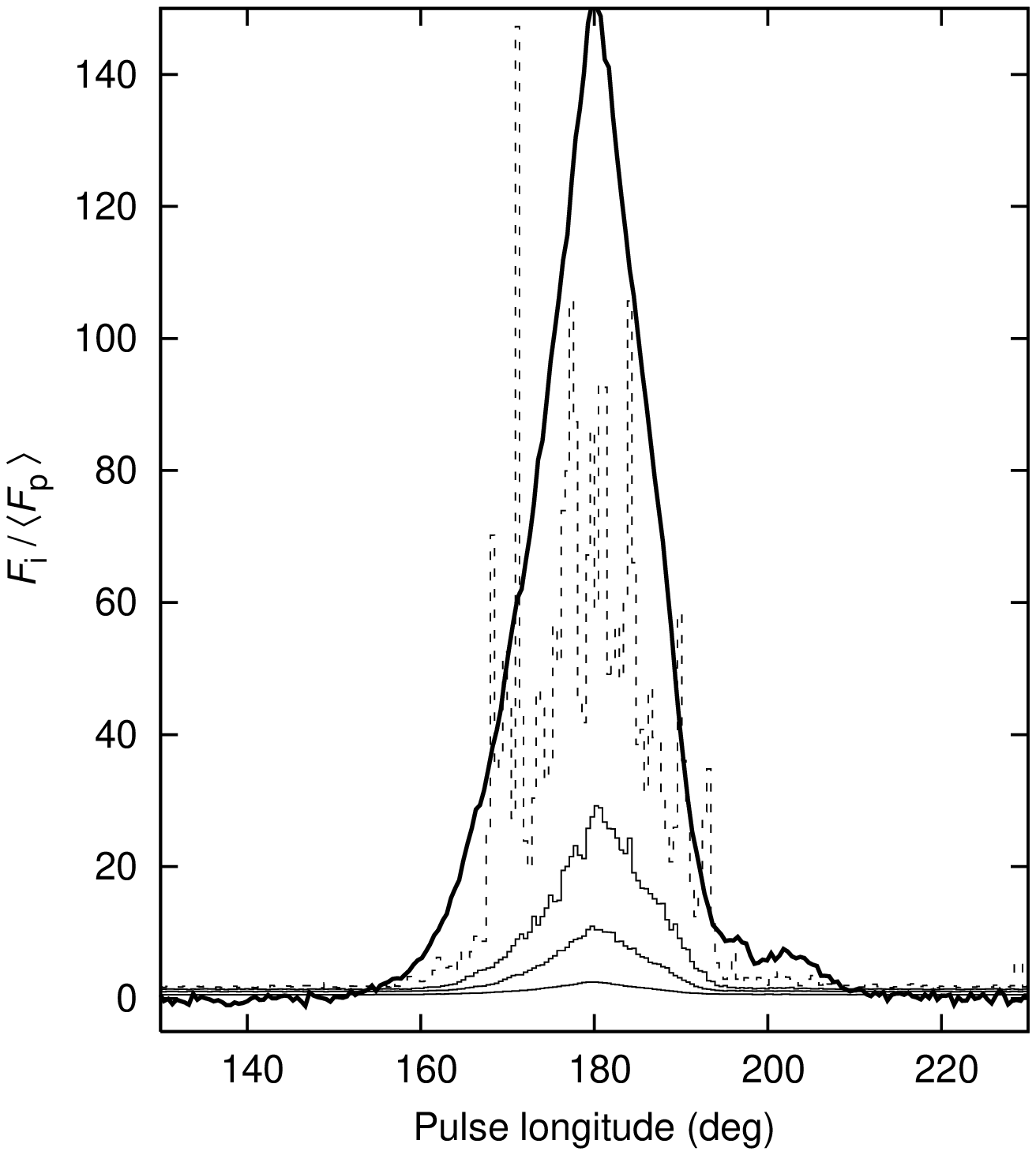}}
\resizebox{!}{0.39\hsize}{\includegraphics[angle=0,trim=0 0 0 0,clip=true]{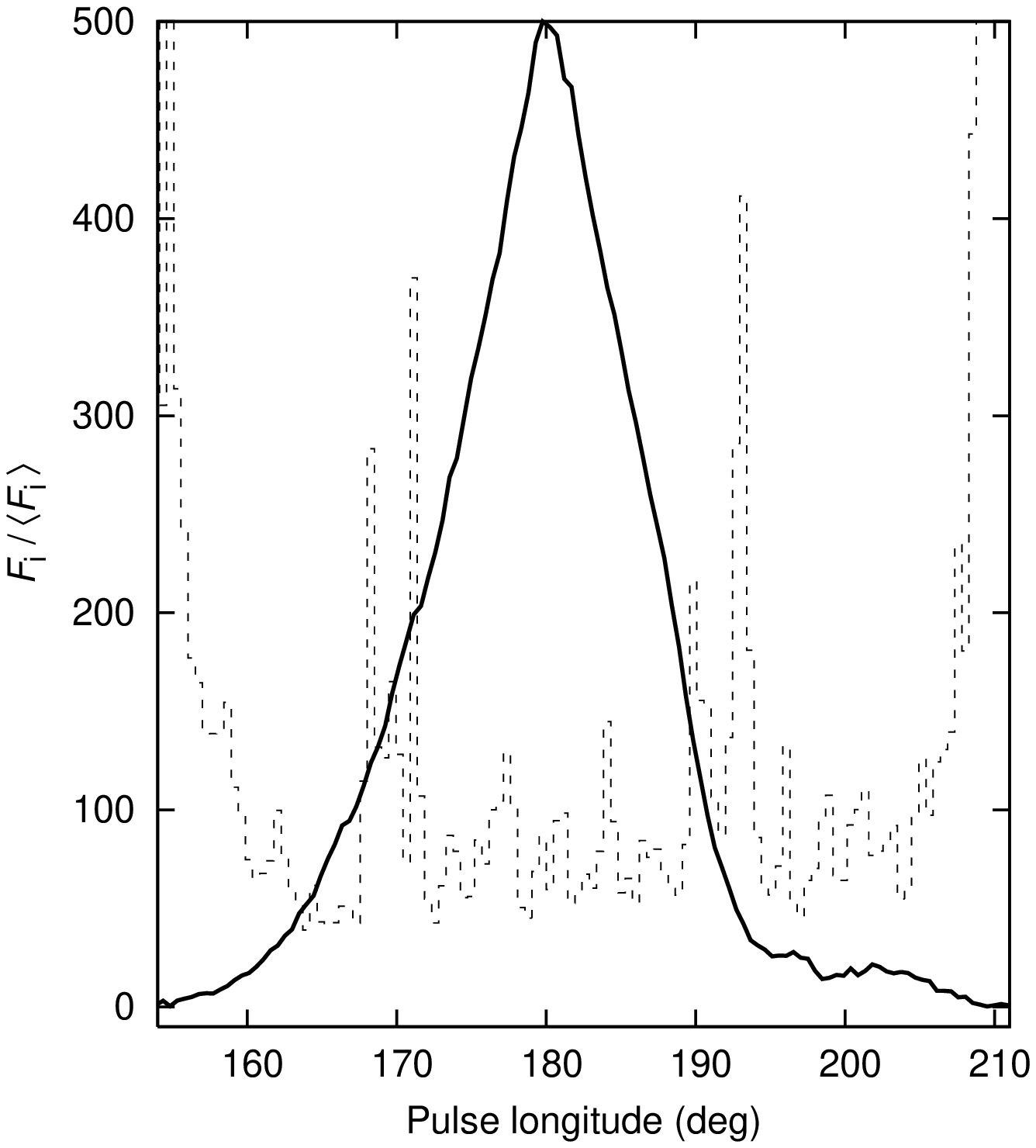}}
\end{center}
\caption{\label{lrced}The top and bottom panels are from the 327-MHz
AO-P1 and 1525-MHz AO-L observation respectively. {\bf Left:}
\label{ProfileBuildup}The top thick line is
the average-pulse profile. The other profiles are (from top to bottom)
the average profile of the pulses with $E>20$ {\Eav}, 10 -- 20 {\Eav},
5 -- 10 {\Eav} and 0 -- 5 {\Eav} for the 327-MHz observation and
contain 15, 87, 484 and 19926 pulses respectively. For the 1525-MHz
observation these profiles are for the pulses with an integrated
energy $E>5$ {\Eav}, 2.5 -- 5 {\Eav} and 0 -- 2.5 {\Eav} and contain
39, 462 and 11426 pulses respectively. All the profiles are normalized
and given a vertical offset for clarity. {\bf Middle:} The top thick
solid line is the pulse profile. The other lines are contour levels of
the longitude-resolved cumulative-energy distribution. The dotted line
shows the brightest time sample for each pulse-longitude bin (compared
with the average peak flux {\Fp} at the pulse longitude of the peak of
the pulse profile). The top solid contour level shows the energy of
the 10th but brightest sample for each pulse longitude. The others
show the energy levels of the cumulative distribution at (from top to
bottom) the 0.1\%, 1\% and 10\% level. {\bf Right: } The solid line is
the pulse profile. The dotted line shows the brightest time sample for
each pulse-longitude bin compared with the average-pulse intensity
{\Fi} at that pulse longitude.}
\end{figure*}

Another phenomenon, possibly related to giant pulses, are giant
micropulses (\citealt{jvkb01}). These are pulses that are not
necessarily extreme in the sense that they have a large integrated
pulse energy $E$, but their peak flux densities are very large. To
investigate this phenomenon the longitude-resolved cumulative-energy
distributions are calculated and the contour levels are shown in the
middle panels of Fig. \ref{lrced}. In these plots the 1\% contour
shows the flux density that 1\% of pulses exceed. So at a
pulse-longitude bin $i$ the 1\% contour shows that 1\% of all the
pulses have (in that single longitude bin $i$) a flux density $F_i$
that is at least as bright as indicated by the contour. The flux
density is compared with the average flux density {\Fp} at the pulse
longitude of the peak of the pulse profile.

One can see that the highest measured peak flux of a single pulse in
the 327-MHz AO-P1 observation is almost 400 {\Fp}.  In the AO-P2
observation the brightest single pulse (which is also located in the
centre of the pulse profile) has a peak flux of $\sim$420 {\Fp}. This
is an order of magnitude brighter than the giant micropulses observed
for the Vela pulsar (\citealt{jvkb01}) and PSR B1706--44 \citep{jr02}.

The giant micropulses of the Vela pulsar are observed to have nearly
complete linear polarization (\citealt{jvkb01}) and are limited to the
leading edge of the pulse profile. The bright pulse in
Fig. \ref{polarization_plot} (with an integrated energy of 22.9
{\Eav}) is almost 100\% linearly polarized, resulting in a linear
polarization profile that is almost indistinguishable from the total
intensity profile. However, in the case of PSR B0656+14 the ``giant
micropulses'' are less confined. Although, like the brightest
integrated pulses, the pulses with the largest peak fluxes are limited
to the leading and central regions and are not found in the shoulders
at the trailing side of the pulse profile.

\begin{figure*}[tb]
\begin{center}
\resizebox{!}{0.295\hsize}{\includegraphics[angle=0,trim=0 0 0 0,clip=true]{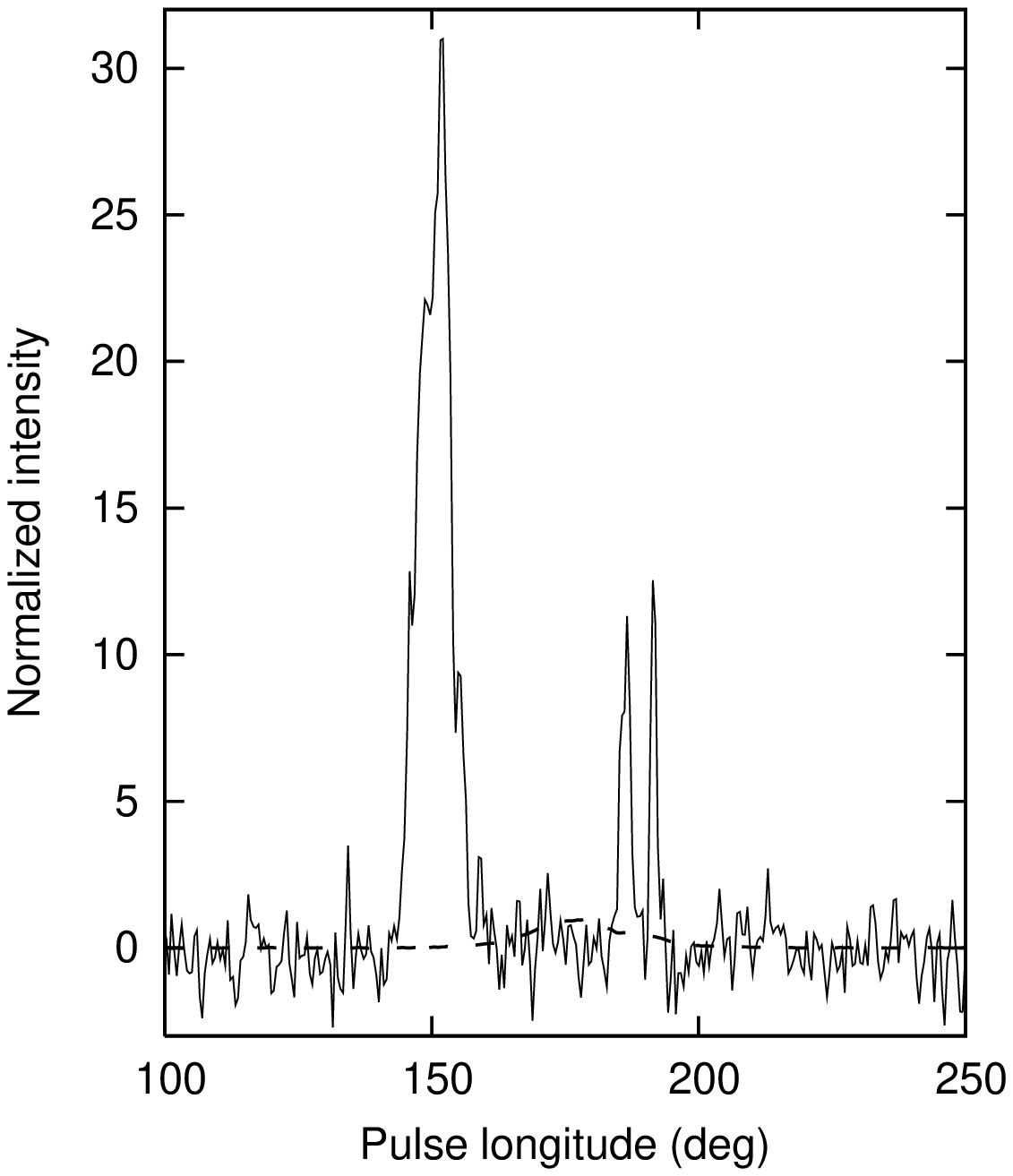}}
\resizebox{!}{0.295\hsize}{\includegraphics[angle=0,trim=0 -10 0 0,clip=true]{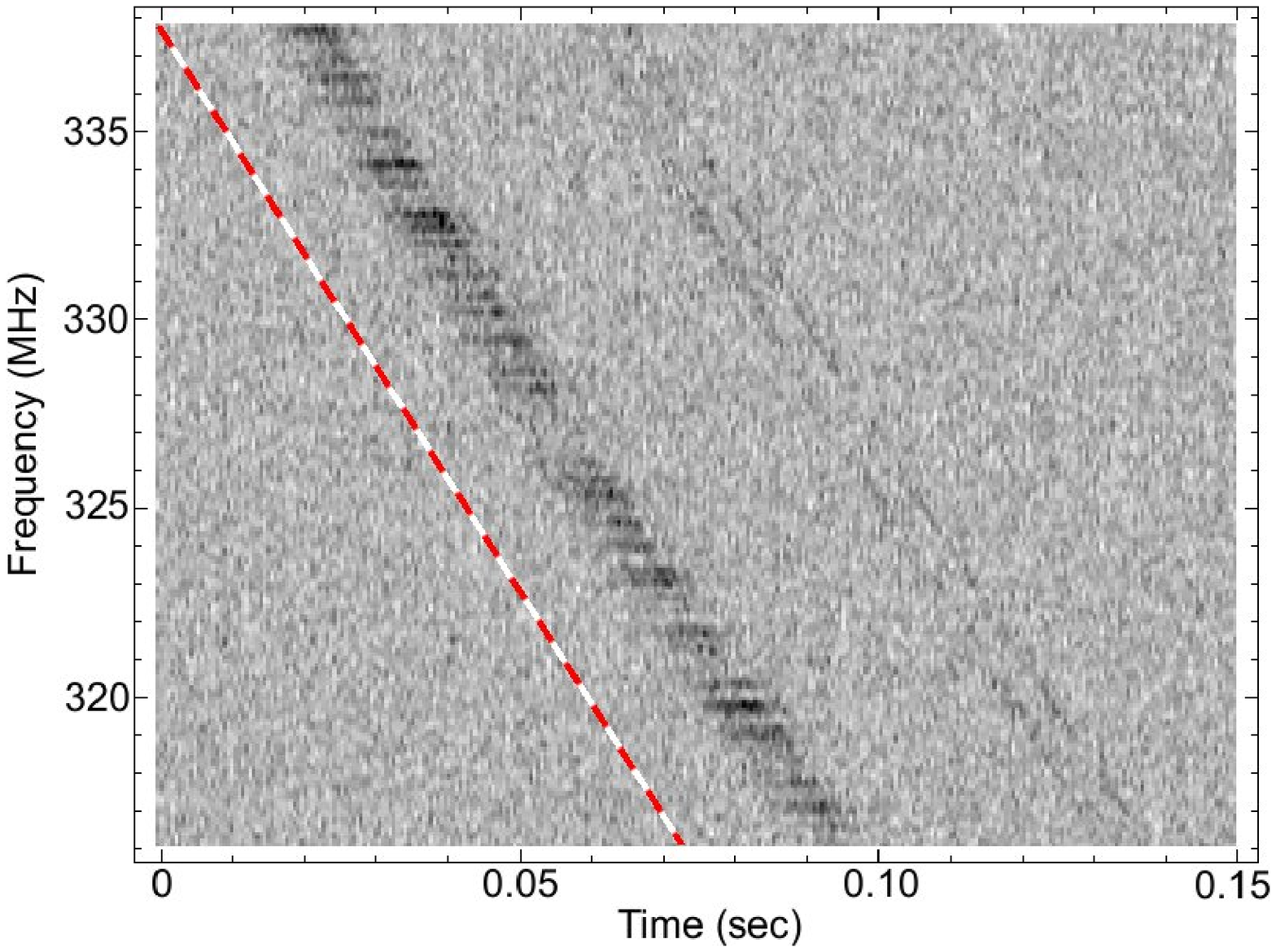}}
\hspace{1mm}
\resizebox{!}{0.295\hsize}{\includegraphics[angle=0,trim=0 0 0 0,clip=true]{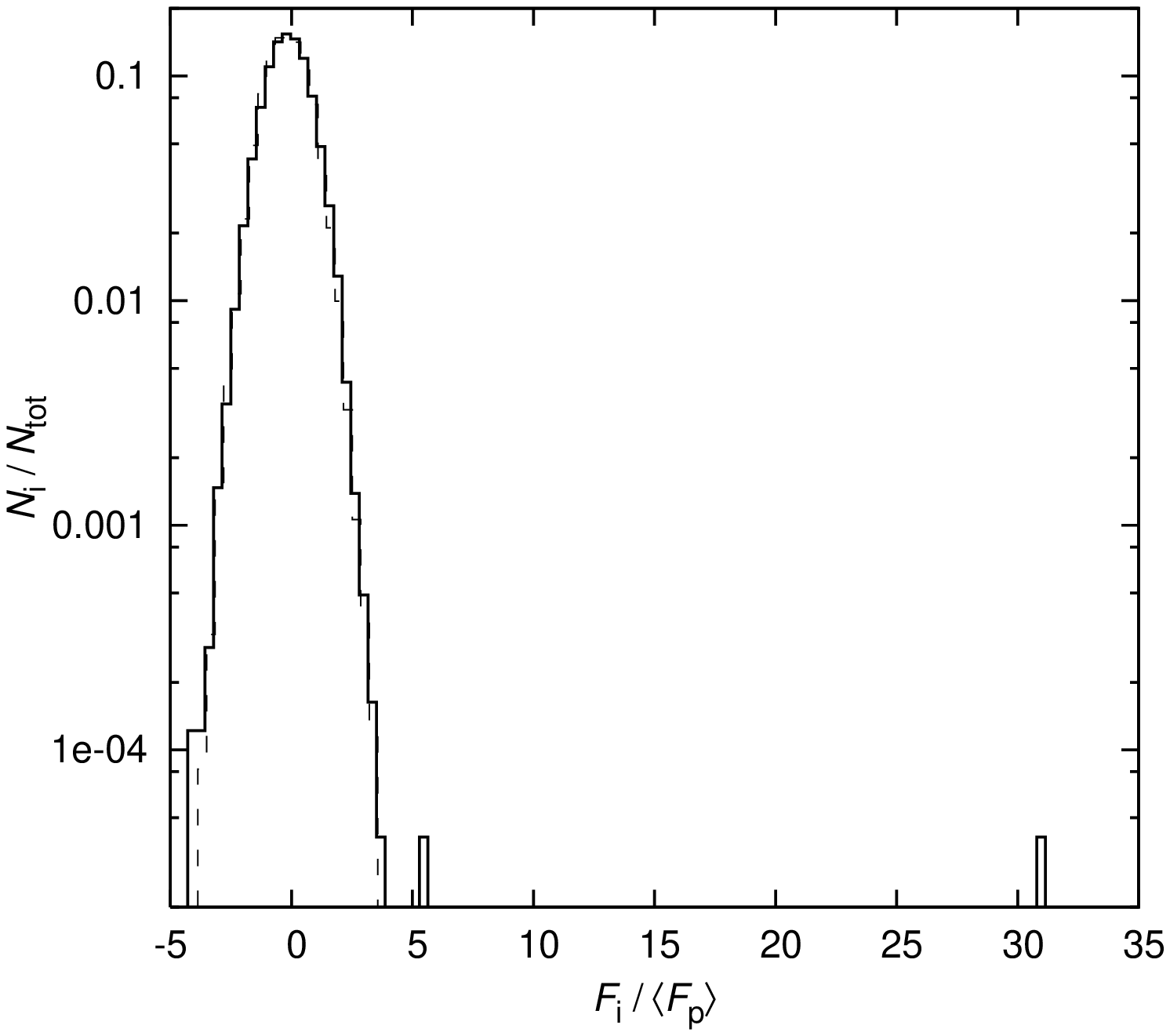}}
\end{center}
\caption{\label{megapulse}The exceptional pulse at the leading edge of
the pulse profile in the the AO-P1 observation.  {\bf Left:} The
single pulse (solid line) compared with the average of all the pulses
(dashed line). The fluxes are normalized such that the peak of the
average pulse is 1.  {\bf Middle:} \label{megapulse_disp} The same
pulse, but now with frequency resolution (non-dedispersed). The dashed
line shows the expected dispersion track for the DM of this
pulsar. {\bf Right: }\label{megapulse_endist}The longitude-resolved
energy distribution at the longitude of the peak of the strong pulse
(solid line) and the off-pulse distribution (dashed line).}
\end{figure*}

Yet the bright pulses found in the centre of the profile are not, in
relative terms, the most extreme examples of spikes in the
emission. For instance at pulse longitude \degrees{157} there is a
single pulse that has a peak flux of 32 {\Fp} (see the top middle
panel of Fig. \ref{lrced}). This is not unusual for the bright pulses
in central regions of the profile. However, the peak flux of this
pulse is almost 1700 times stronger than the average-pulse intensity
{\Fi} at that pulse longitude (see right panel of
Fig. \ref{lrced}). At the outer edges of the pulse profile the
uncertainty in the average-pulse intensity becomes large, making the
uncertainties in the relative peak flux on the edges of the right
panels of Fig. \ref{lrced} very large. So for instance the huge peak
at pulse longitude \degrees{218} in the top right panel of
Fig. \ref{lrced} is probably caused by a statistical fluctuation that
makes {\Fi} small, rather than by a bright single pulse. Note that the
bright pulse at pulse longitude \degrees{157} is both seen as a spike
in the dashed line of the middle panel of Fig. \ref{lrced} as well as
an increase in intensity of the pulse profile.

\subsection{\label{SctStronPulse}An exceptional pulse}

The most exceptional bright pulse is located on the leading edge of
the pulse profile (the bump in the dashed line starting at pulse
longitude \degrees{143} in the middle top panel of Fig. \ref{lrced}).
This single pulse is plotted in the left panel of
Fig. \ref{megapulse}. With an integrated pulse energy of 12.5 {\Eav}
the pulse is very strong, but not unusual for this pulsar (although it
qualifies as a giant pulse). What makes this pulse special is that the
pulse is extremely bright compared with the local average-pulse
profile intensity. Its peak flux is about 2000 times that of the
average-pulse intensity at that location.

Especially because this pulse is found well away from the centre of
the profile, it is important to check if this pulse is not generated
by RFI. To prove that this pulse is indeed emitted by the pulsar, the
pulse is plotted with frequency resolution (middle panel of
Fig. \ref{megapulse}) and is not de-dispersed. A non-dispersed (Earth
related) signal would be visible as a vertical band in this plot,
however the dispersion track matches what is expected from the DM of
this pulsar (and is the same for the two pulses in the centre of the
profile) proving that this exceptional pulse is emitted by the pulsar.
Notice also that the effect of scintillation is clearly visible (which
is further evidence of the reality of the signal) and that the
scintillation bandwidth is much smaller than the bandwidth of the
observation (as calculated in Sect. \ref{scintct}). Notice also that
the scintillation pattern of the bright pulse is the same as that of
the two peaks in the middle of the pulse profile.

To emphasize how extreme this pulse is, the longitude-resolved energy
distribution at the position of the peak of this pulse is
calculated (right panel of Fig. \ref{megapulse_endist}). The peak flux
of the pulse is 31 times higher than the average peak flux of the
pulse profile. There is only one other detection of a single pulse at
the same pulse longitude (with a flux density of 5.5 {\Fp}). The
remaining 25,000 pulses do not have intensities above the noise level.
This implies that these two pulses either belong to an extremely long
tail of the energy distribution, or that only very sporadic
pulses are emitted at this pulse longitude.

\subsection{The pulse-energy distribution}

The pulse-energy distributions of giant pulses can be described by a
power law (e.g. \citealt{lcu+95}), while the pulse-energy distribution
of ``normal'' pulses can often be described by lognormal statistics
(e.g. \citealt{cai04,cjd01,jvkb01}).  Because only a fraction of the
pulses are giant pulses, the energy distribution of all the pulses
consist of two components and a break might appear. From
Fig. \ref{Enhists} it is clear that there is no obvious break in the
pulse-energy distribution of PSR B0656+14.  While no clear breaks are
observed in the pulse-energy distributions, a break could be hidden
below the noise level. Therefore an important question we should now
ask is whether the whole energy distribution (including the part that
is below the noise level) can be described by a single distribution,
or is a multi-component distribution.

The very bright nature of some of the pulses from PSR B0656+14 may
make a contribution to the overall noise received by the telescope and
thus the off-pulse noise level would not be an accurate estimator of
the on-pulse noise. However, we find that the pulsar signal is
sufficiently smeared in time by interstellar dispersion to only
increase the overall noise level with a few percent during the
brightest recorded pulse. Moreover, the majority of the pulses are
much weaker than this particular pulse, so it is appropriate to
continue to use the off-pulse noise when estimating the $S/N$ of the
pulses.

\begin{figure*}[htb]
\begin{center}
\resizebox{0.49\hsize}{!}{\includegraphics[angle=0,trim=-13 0 0 0,clip=true]{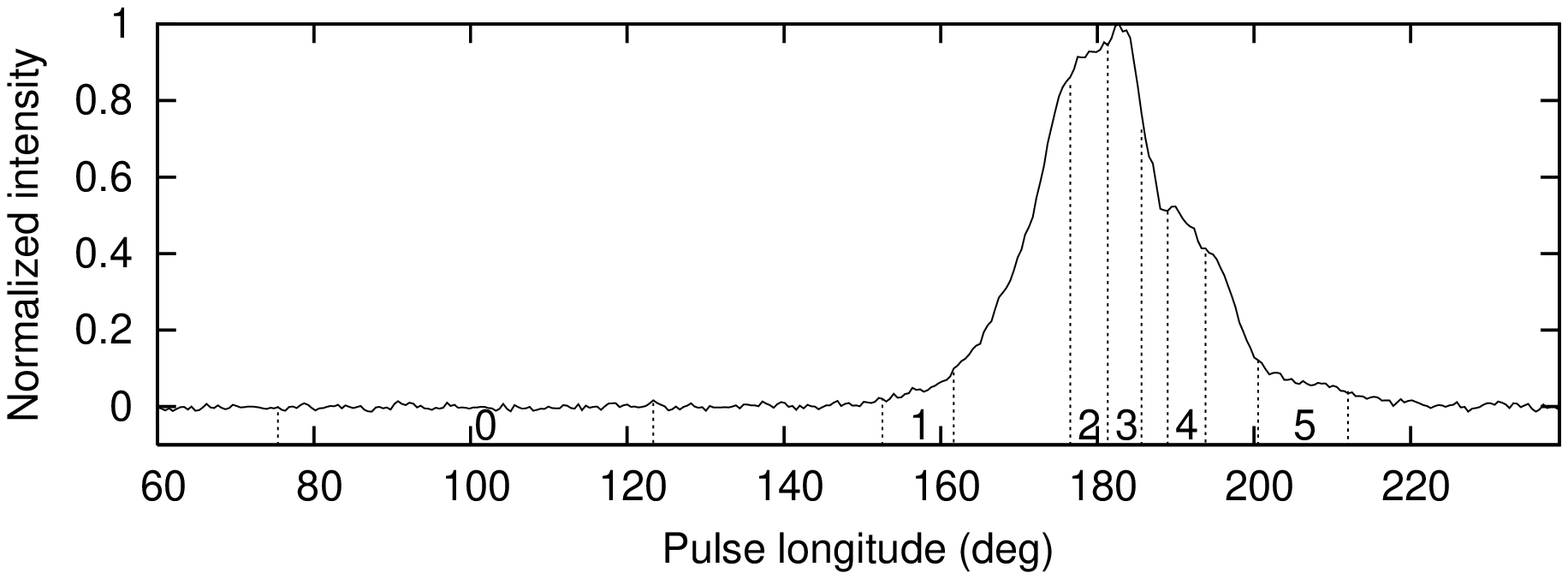}}
\hspace{0.01\hsize}
\resizebox{0.49\hsize}{!}{\includegraphics[angle=0,trim=-13 0 0 0,clip=true]{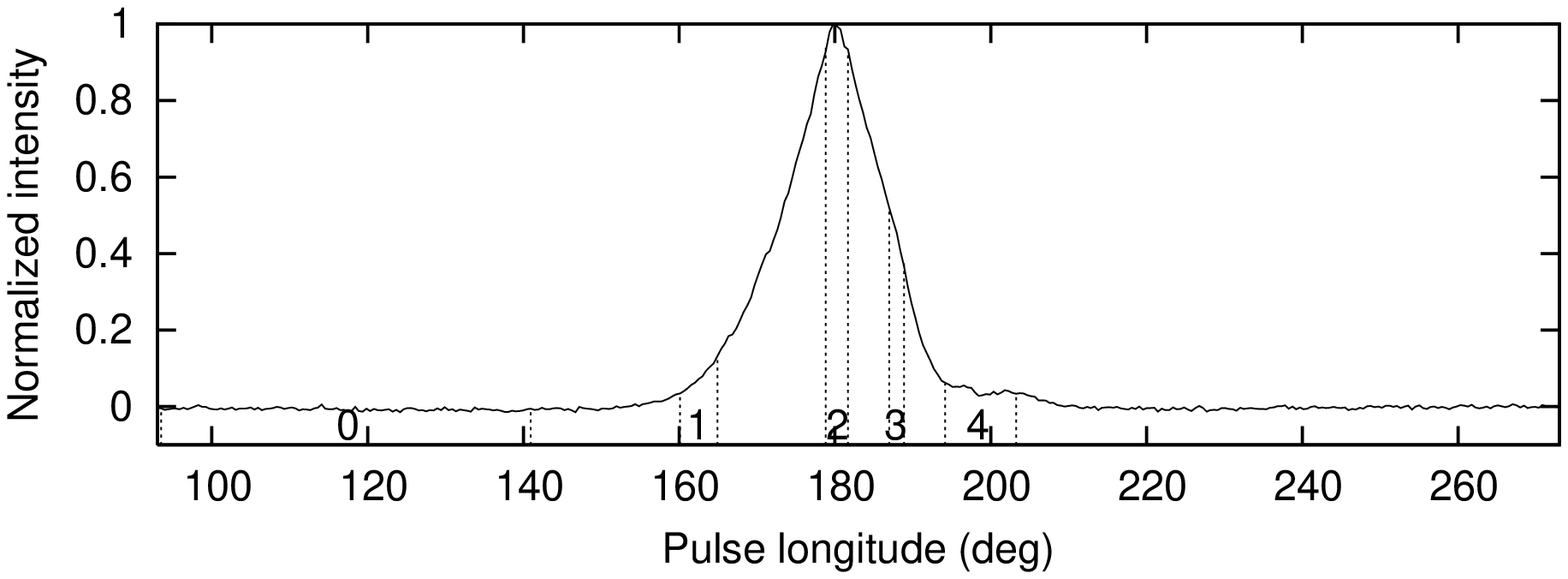}}\\
\vspace*{2mm}
\resizebox{0.49\hsize}{!}{\includegraphics[angle=0,trim=0 0 0 0,clip=true]{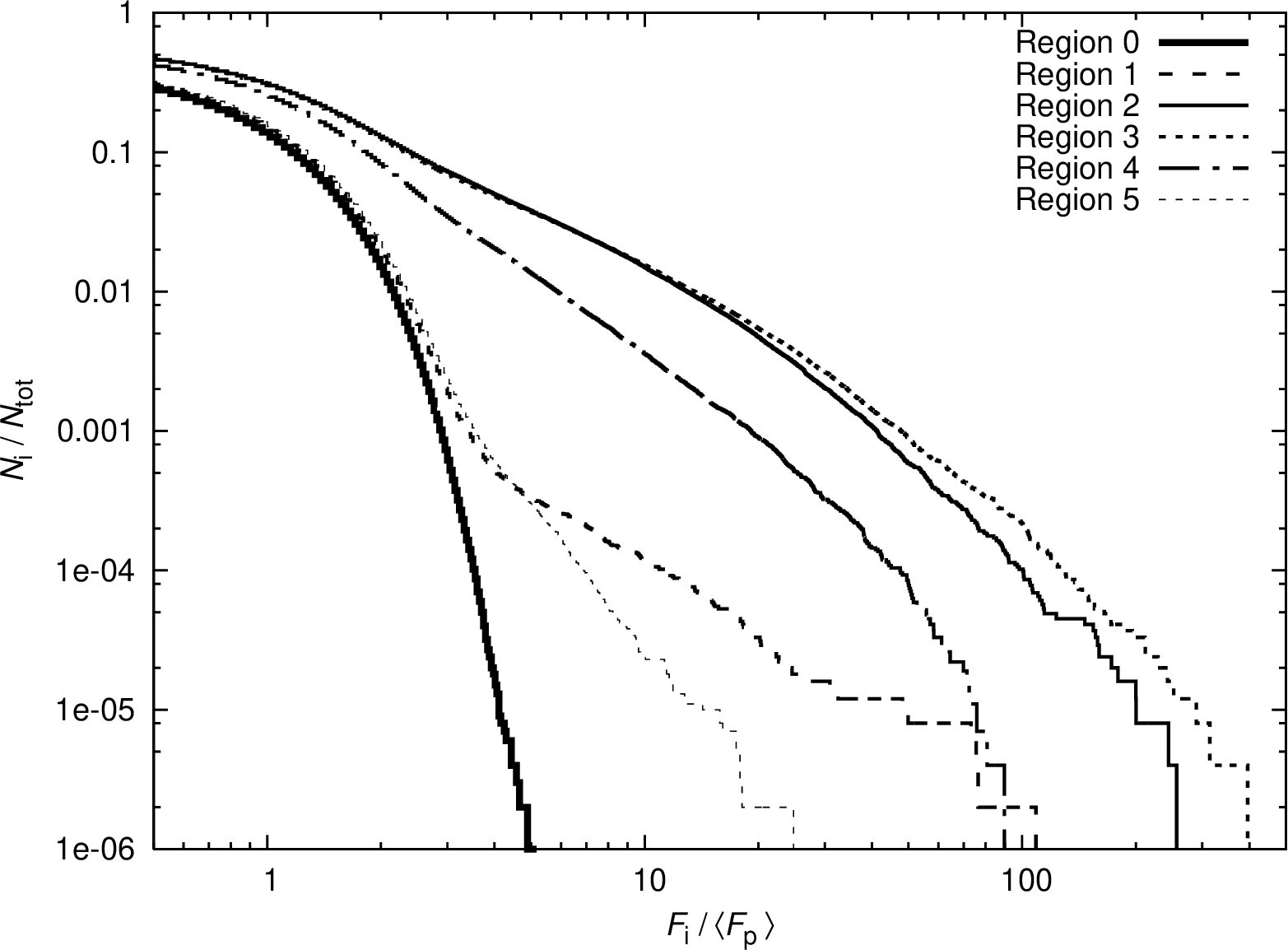}}
\hspace{0.01\hsize}
\resizebox{0.49\hsize}{!}{\includegraphics[angle=0,trim=0 0 0 0,clip=true]{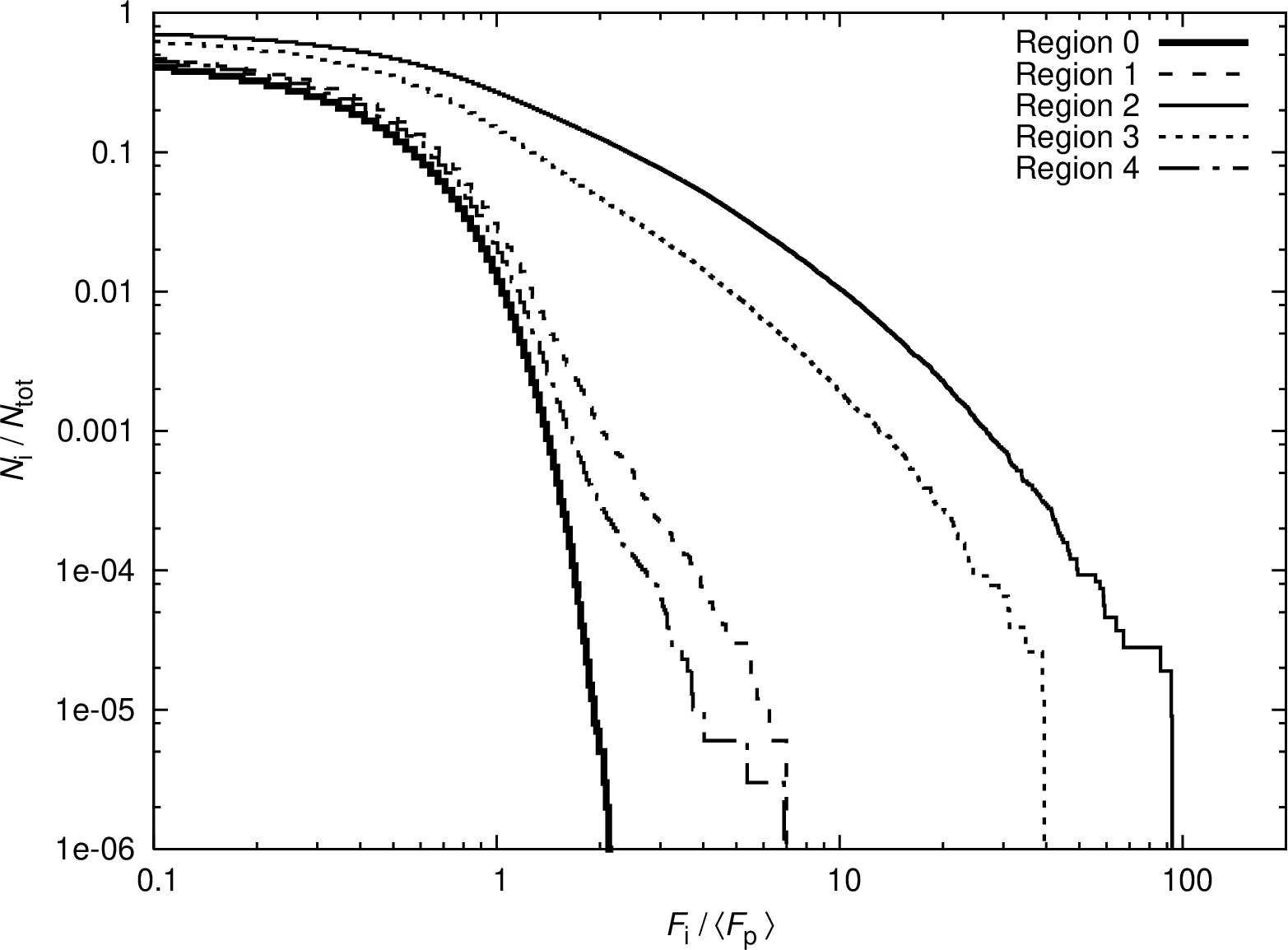}}
\end{center}
\caption{\label{lrced_regions}The longitude-resolved cumulative-energy
distribution for different regions (between the lines) in the pulse
profile. The left and right panels are for the 327-MHz AO-P1 and the
1525-MHz AO-L observations respectively.  }
\end{figure*}

\subsubsection{Fitting procedure of the pulse-energy distribution}

The pulse-energy distribution is modeled with a single lognormal or a
power law.
\begin{eqnarray}
\nonumber P_\mathrm{powerlaw}(E)&\propto& E^p\\
P_\mathrm{lognormal}(E)&=&\frac{<\!E\!>}{\sqrt{2\pi}\sigma E}\exp\left[-\left(\ln \frac{E}{<\!E\!>}-\mu\right)^2/\left(2\sigma^2\right)\right].
\end{eqnarray}
Because the integral of a power-law distribution is infinite, a cutoff
energy $E_\mathrm{min}$ (the minimum energy of the pulses) must be
introduced. So the power-law distribution, as well as the lognormal
distribution, are defined by two fit parameters: $p$ and
$E_\mathrm{min}$ in the case of a power-law distribution and $\mu$ and
$\sigma$ in the case of a lognormal distribution.

The possible existence of two distributions, one of strong and one of
weak pulses, is modeled by adding pulses with zero energy to the model
distribution. It is important to note that these pulses with zero
energy are not necessarily representing real ``nulls''. They represent
a distribution of pulses that are significantly weaker than the tail
of strong pulses. This distribution of weak pulses can be described by
only a single parameter: the fraction of weak pulses
($f_\mathrm{weak}=N_\mathrm{weak}/N_\mathrm{tot}$).

The three model parameters are not independent of each other. They are
coupled by the requirement that the average-pulse energy {\Eav} of the
model distribution should match the observed value. This constraint is
used to fix the fraction of weak pulses for a given combination of the
other two fit parameters. Although this description of the
pulse-energy distribution is probably oversimplified, it has the
advantage that there are only two fit parameters. Given the $S/N$
ratio of the observations, the addition of more fit parameters
would probably over-interpret the available information.

To take into account the effect of noise, the model energy
distribution is convolved with the noise distribution.  Because of the
presence of some RFI, the noise distributions of the AO observations
deviate from a pure Gaussian distribution. Therefore the observed
noise distribution is used rather than a theoretical noise
distribution. For a given set of fit parameters, the model
pulse-energy distribution is compared with the observed
distribution. The fit parameters are optimised by minimizing the
$\chi^2$ using the downhill simplex method (amoeba algorithm;
\citealt{pftv86}).  Fitting is restricted to bins with more than 10
counts and counting statistics are used for the measurement
uncertainties in each bin.  The fitting depends on the binning used,
and therefore we checked that all results are valid for different
binning and count limits.

\subsubsection{Results of fitting the pulse-energy distributions}

\begin{table}
\caption{\label{PowerlawIndex}The parameters of the best fits to the
pulse-energy distributions of the four observations. Besides the
parameters of the fitted distribution (where
$f_\mathrm{weak}=N_\mathrm{weak}/N_\mathrm{tot}$ is the fraction of
weak pulses that are added to the distribution), also the total
$\chi^2$, the number of degrees of freedom and the significance
probability are tabulated.}
\begin{center}
\begin{tabular}{r@{-}l|ccc|ccr@{$\times$}l}
\hline
\hline
\multicolumn{2}{c|}{\hspace*{3.5mm}REF} & $\sigma$ & $\mu$ & $f_\mathrm{weak}$ & $\chi^2$ & $N_{DOF}$ & \multicolumn{2}{c}{$P(\chi^2)$}\\
\hline
AO&P1 & 0.99 & $-0.34$ & 13\% & 105 & 32 & 1&$10^{-9}$\\
AO&P2 & 0.79 & $-0.34$ & 0\% & 146 & 24 & 2&$10^{-19}$\\
AO&L  & $0.56$ & $-0.13$ & 0\% & 240 & 44 & 8&$10^{-29}$\\
WSRT&L  & $0.48$ & $-0.03$ & 8\% & 148 & 44 & 4&$10^{-13}$\\
\hline
\end{tabular}
\end{center}
\end{table}

For all four observations we fit a cutoff power law and a lognormal
distribution (convolved with the observed noise distribution) to the
pulse-energy distribution. As described above we include the
possibility for a separate distribution of weak pulses. In all cases
the fits using a cutoff power law are worse than the fits with a
lognormal distribution. The results of the fits with a lognormal
distribution are summarized in Table \ref{PowerlawIndex} and in
Fig. \ref{Enhists} the fits (convolved with the noise distributions)
are shown overlaid over the observed distributions. As one can see the
fits can qualitatively describe the observations well.

The significance of the best fits are low, indicating that the model
is oversimplified. Furthermore, the best fits for the different
observations at the same frequency are quite different. For instance
the fit to the AO-P1 data is significantly improved by adding a
distribution of weak pulses while the addition of weak pulses to the
AO-P2 data does not improve the fit. This probably indicates that the
observations are too short to get a pulse-energy distribution that is
stable. This may be related to the gradual profile change described in
Sect. \ref{SctStability} as the fits for the first and second halves
of the AO-P2 observation are different. The fit for the first half of
the observation (when the central peak is relatively bright) is
improved by adding a distribution of weak pulses, while this is not
the case for the second half of the observation.

The pulse-energy distribution of PSR B0656+14 can (at least for some
time intervals) only be described by a multicomponent distribution
consisting of a distribution of strong pulses and weak pulses. The
absence of a clear break in the observed distribution may indicate
that the transition from weak to strong pulses is smooth. If a sharp
break exists, it must be hidden below the noise.

\subsubsection{The longitude-resolved energy distribution}

In Fig. \ref{lrced_regions} examples of longitude-resolved
cumulative-energy histograms are shown. These are the energy
histograms of single pulse-longitude bins. For better statistics,
energy histograms of multiple successive pulse-longitude bins in a
small longitude range are added. The regions used are shown in
Fig. \ref{lrced_regions} as well.  It is interesting to see the
difference in the cumulative distribution of region 1 and 5 of the
327-MHz AO-P1 observation. Although the regions are comparable in
intensity in the average-pulse profile, the leading side of the
profile shows much brighter spikes. As for for the pulse-energy
distributions, the longitude-resolved energy distributions do not show
a clear break. The apparent break in the longitude-resolved energy
distributions at the cutoff energy of the noise distribution (best
visible for region 1 in the bottom left panel of
Fig. \ref{lrced_regions}) is most likely not a physical break in the
energy distribution, but just the transition from noise-dominated to a
pulse-dominated energy range.

\begin{figure}[tb]
\begin{center}
\resizebox{0.99\hsize}{!}{\includegraphics[angle=270,trim=0 0 0 0,clip=true]{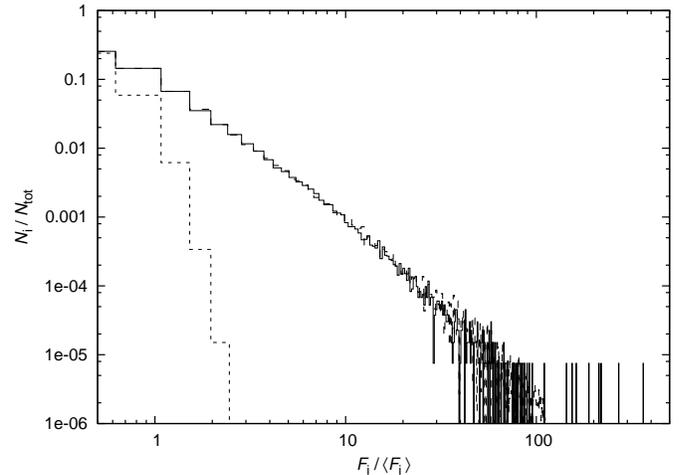}}
\end{center}
\caption{\label{lred_fit}The tail of the longitude-resolved energy
distribution of the peak of the 327-MHz AO-P2 observation (solid
line), the fit (dashed line) and the off-pulse distribution (dotted
line).  }
\end{figure}

Using exactly the same method as used to fit the pulse-energy
distribution we tried to fit the longitude-resolved energy
distributions. Again it turns out that a cutoff power law cannot
describe the data well, but a lognormal distribution can be fit with
much more confidence. For instance, the longitude-resolved energy
distribution of the peak of the AO-P2 observation can be described
well with a lognormal distribution ($\mu=-1.5$ and
$\sigma_\mathrm{lognorm}=1.7$, without a distribution of weak pulses).
As one can see in Fig. \ref{lred_fit}, the fit describes the data
qualitatively very well. The total $\chi^2=104$ with $61$ degrees of
freedom ($P(\chi^2)=5\times10^{-4}$).  This is a better fit than those
found for the integrated pulse-energy distribution, which is
surprising because the longitude-resolved energy distribution has
better statistics because of two reasons. Firstly, the distributions
of 7 bins were added which makes the counting uncertainties
smaller. Secondly, the longitude-resolved distribution is much more
extended (compared with the off-pulse distribution) than the
integrated pulse-energy distribution.

\subsubsection{Two populations of pulses?}

There is no direct evidence for the existence of a separate energy
distribution for the strongest pulses of PSR B0656+14 as there is no
clear break in the pulse-energy distribution. Nevertheless there is
evidence from the energy distributions that the bursts and the
underlying weak emission have different properties.

The first argument is based on the frequency evolution of the
pulse-energy distribution. Although the exact shape of the
pulse-energy distribution remains unclear, it is steeper and less
extended at high frequencies ($\sigma$ is smaller). The difference in
shape of the distributions at the two frequencies is not only a
scaling in energy range. The energy distribution of the high frequency
observation is too steep to be scaled to fit the 327-MHz observations,
which could be because the emission shows more bright pulses at low
frequencies.

\begin{figure}[tb]
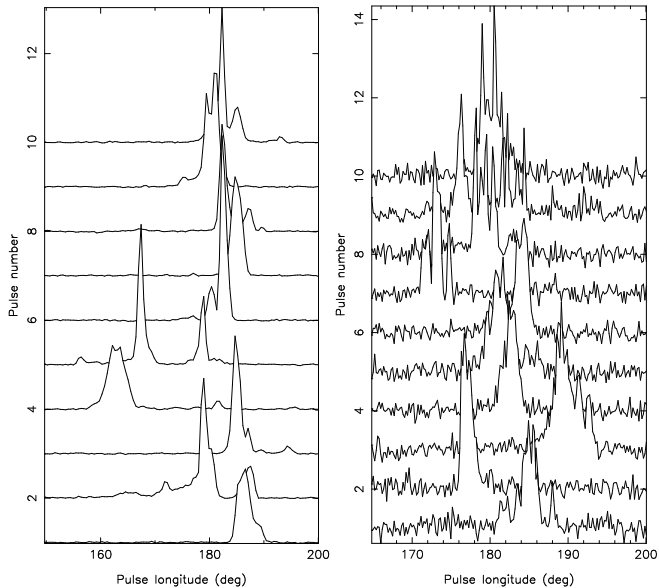

\begin{center}
\resizebox{0.485\hsize}{!}{\includegraphics[angle=0]{5572f11a.ps}}
\resizebox{0.485\hsize}{!}{\includegraphics[angle=0]{5572f11b.ps}}
\end{center}
\caption{\label{BrightestPulses}The ten brightest pulses of the
327-MHz AO-P1 observation (left panel) and of the 1380-MHz WSRT-L
observation (right panel).}
\end{figure}

Secondly, as discussed above, the longitude-resolved distribution is
fitted much better by a lognormal distribution than the integrated
pulse-energy distribution. This is probably because when a single
pulse contains a very bright subpulse, it is more likely to contain
more bright subpulses (see for example
Fig. \ref{BrightestPulses}). This affects only the integrated
energy. This is therefore evidence that the bursts of radio emission
not only cluster in successive pulses, but they also tend to cluster
within the pulses. This clustering generates a tail in the
pulse-energy distribution which is relatively more extended than that
of the longitude-resolved energy distribution. This also explains why
all observations show a flattening in their integrated pulse-energy
distribution (containing only a few pulses) at the highest energies
that cannot be fit (see Fig. \ref{Enhists}).

Thus we have to accept the perhaps surprising conclusion
that, unlike the energy distributions of pulsars with true ``giant''
pulses, those of PSR B0656+14 offer no direct evidence of comprising
two components.  Although it is possible that the bright and weak
pulses are associated with the extreme ends of a single smooth energy
distribution, we will demonstrate in the next section that they have
very different characteristics.

\section{\label{SctSeparation5}Characteristics of the spiky and weak emission}

\subsection{Appearance of the spiky emission}

Perhaps the strongest argument for the presence of a distinctive
``spiky'' emission is the appearance of the pulse stack itself
(Fig. \ref{Stack}). Bright, narrow, subpulses seem to dominate the
emission. This suggests that they may have different properties from
the rest of the emission.  In the left panel of
Fig. \ref{BrightestPulses} the ten brightest pulses of the AO-P1
observation are shown.  One can see that the spikes are not only
narrow, but appear to be quasi-periodic. In the left panel of
Fig. \ref{BrightestPulses} one can see that most strong pulses show a
$\sim$11 ms (\degrees{10}) periodicity (for instance pulse 5). In the
right panel of Fig. \ref{BrightestPulses}, the ten brightest pulses of
the WSRT-L observation are shown, which has a higher time resolution
allowing the detection of structure on shorter timescales. In addition
to the 11-ms periodicity, a 1-ms (\degrees{1}) periodicity is also
revealed (for instance pulse 9). In the rest of this section we will
further characterize the spiky and weak emission in detail.

\subsection{\label{SctSeparation}Separation of the spiky and weak emission}

Although the pulse sequence of Fig. \ref{Stack} is dominated by the
very apparent spiky emission, this is accompanied by an almost
indiscernible background of weak emission.  To separate these two
components of the emission, we have applied an intensity threshold to
the data. The intensities of the time samples in the pulse stack of
the weak emission are truncated if they exceed this threshold. The
time samples in the pulse stack of the spiky emission contains only
samples with intensity in excess of this threshold. When the pulse
stack of the weak emission is added to the pulse stack of the spiky
emission, one retrieves exactly the original pulse stack.

\begin{figure}[tb]
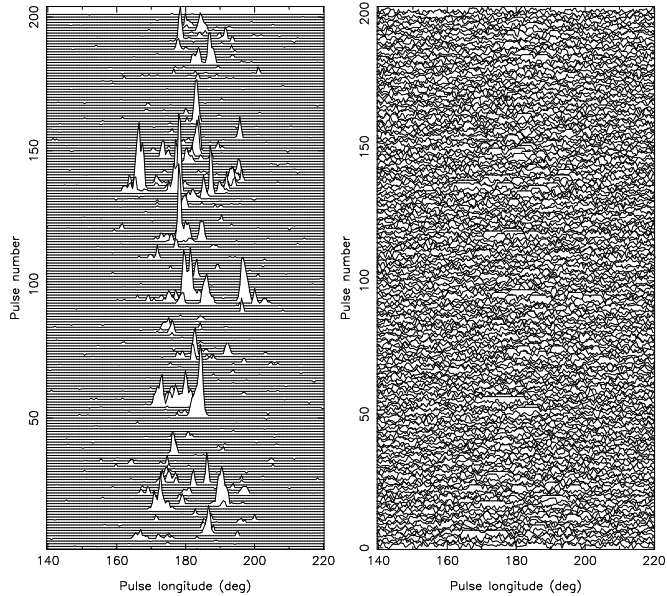

\begin{center}
\resizebox{0.49\hsize}{!}{\includegraphics[angle=0]{5572f12a.ps}}
\resizebox{0.49\hsize}{!}{\includegraphics[angle=0]{5572f12b.ps}}
\end{center}
\caption{\label{StackSep}Here the same 200 successive pulses of the
327-MHz AO-P1 observation as plotted in Fig. \ref{Stack} are shown,
but separated into the spiky (left) and weak (right) emission. }
\end{figure}

Because no clear break is observed in the longitude-resolved energy
distribution (Fig. \ref{lrced_regions}), we have set the threshold
intensity such that 99\% of the noise samples are below this threshold
value.  Only about 1\% of the weak emission is expected to be
classified as spiky emission due to noise fluctuations, so the
emission classified as spiky is expected to be almost pure. Not only
do the noise fluctuations make it impossible to completely separate
the weak and spiky emission, it is also very well possible that the
energy distributions of the two components overlap.  In
Fig. \ref{StackSep} one can see the pulse stack obtained by the
separation of the spiky and weak emission for exactly the same pulse
sequence as shown in Fig. \ref{Stack}. The integrated power of this
sequence of weak pulses is about 3 times greater than that of the
sequence of the spiky emission. This shows that a significant fraction
of the pulsar's emission lies at or below the noise level.

By linking the threshold intensity to the noise level, our analysis
depends on the sensitivity of the telescope. An alternative natural
choice is to stipulate that the integrated weak and spiky emission
profiles have equal power. This resulted in an intensity threshold
very close to the one described above. In general, we found that our
conclusions did not depend on the exact choice of threshold
intensity. This also implies that the effects of interstellar
scintillation in the L-band observations, for which we have made no
correction, will not significantly alter our results.

\begin{figure}[tb]
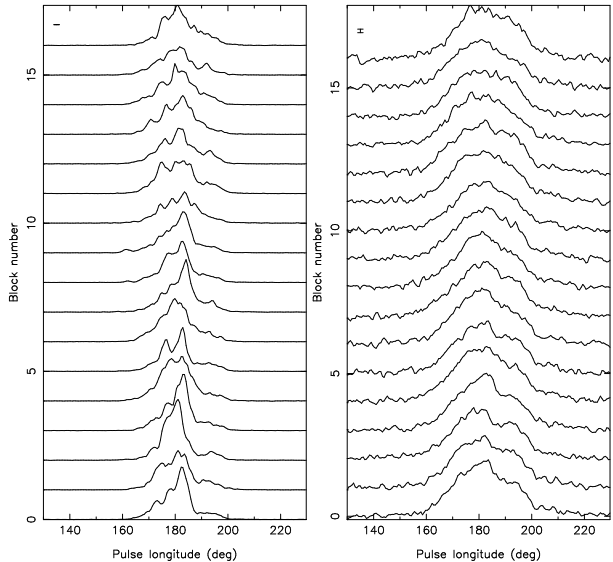

\begin{center}
\resizebox{0.45\hsize}{!}{\includegraphics[angle=0]{5572f13a.ps}}
\resizebox{0.45\hsize}{!}{\includegraphics[angle=0]{5572f13b.ps}}
\end{center}
\caption{\label{ProfilesSep}The pulse profiles obtained by averaging
successive blocks of one thousand pulses each of the 327-MHz AO-P2
observation (like the middle right panels in Fig. \ref{Profiles}), but
now for the spiky and weak emission separately. Note the gradual
changes in the spiky emission profile, in contrast to the steady weak
emission. The 1-sigma error bars are plotted in the top left
corner.}
\end{figure}

\begin{figure*}[htb]
\begin{center}
\resizebox{!}{0.37\hsize}{\includegraphics[angle=0]{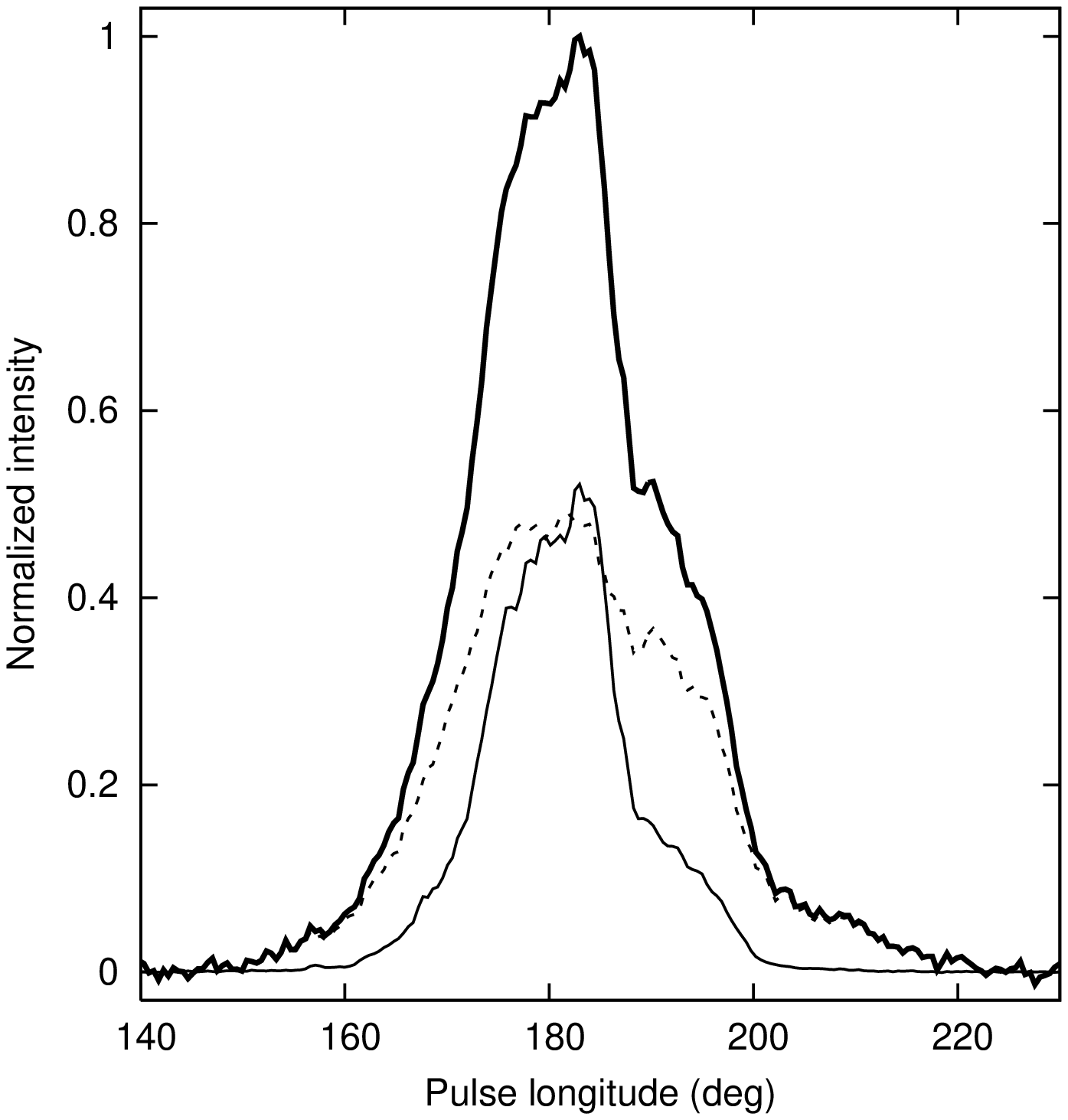}}
\hspace{0.01\hsize}
\resizebox{!}{0.37\hsize}{\includegraphics[angle=0]{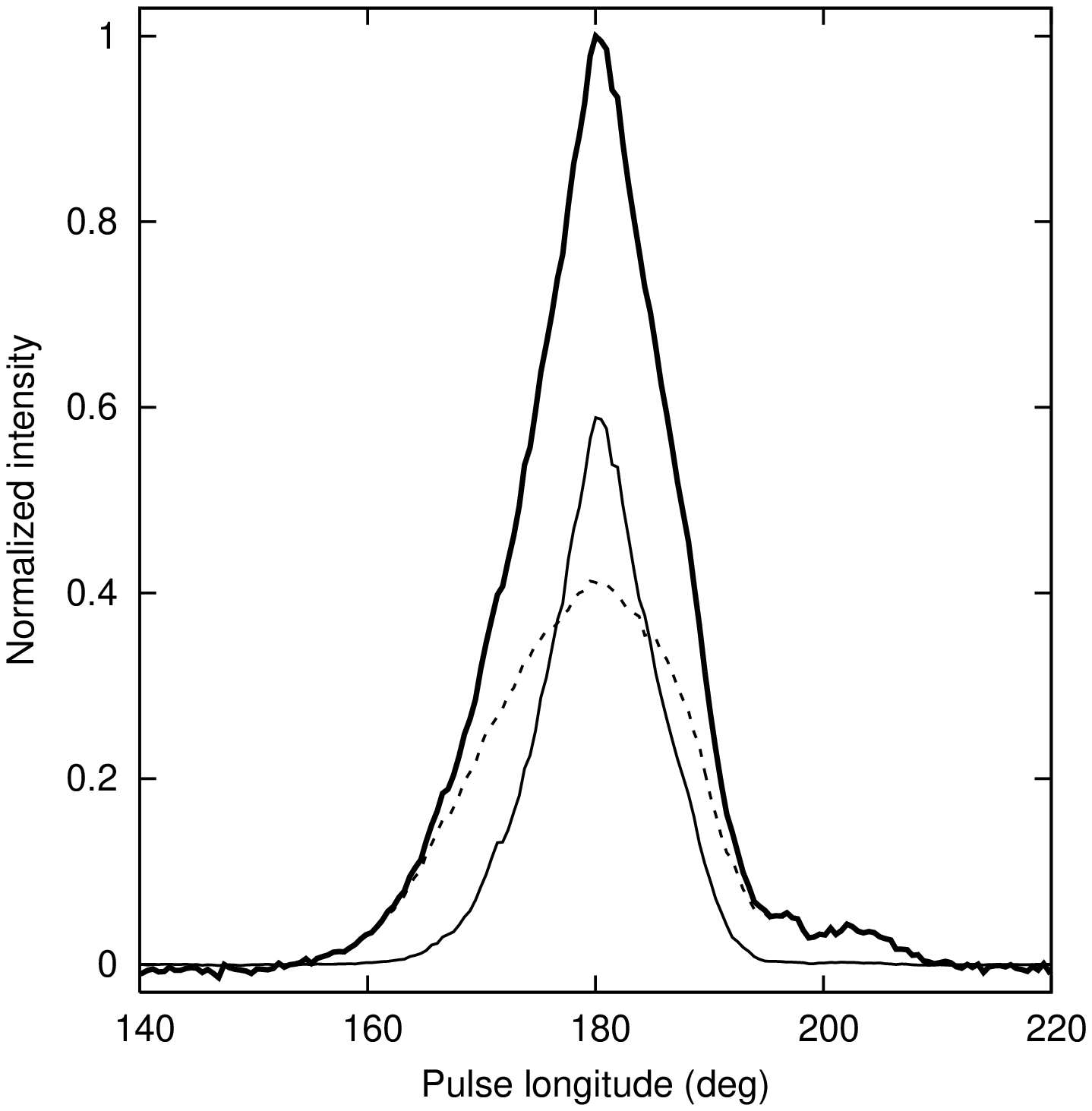}}
\hspace{0.01\hsize}
\resizebox{!}{0.37\hsize}{\includegraphics[angle=0]{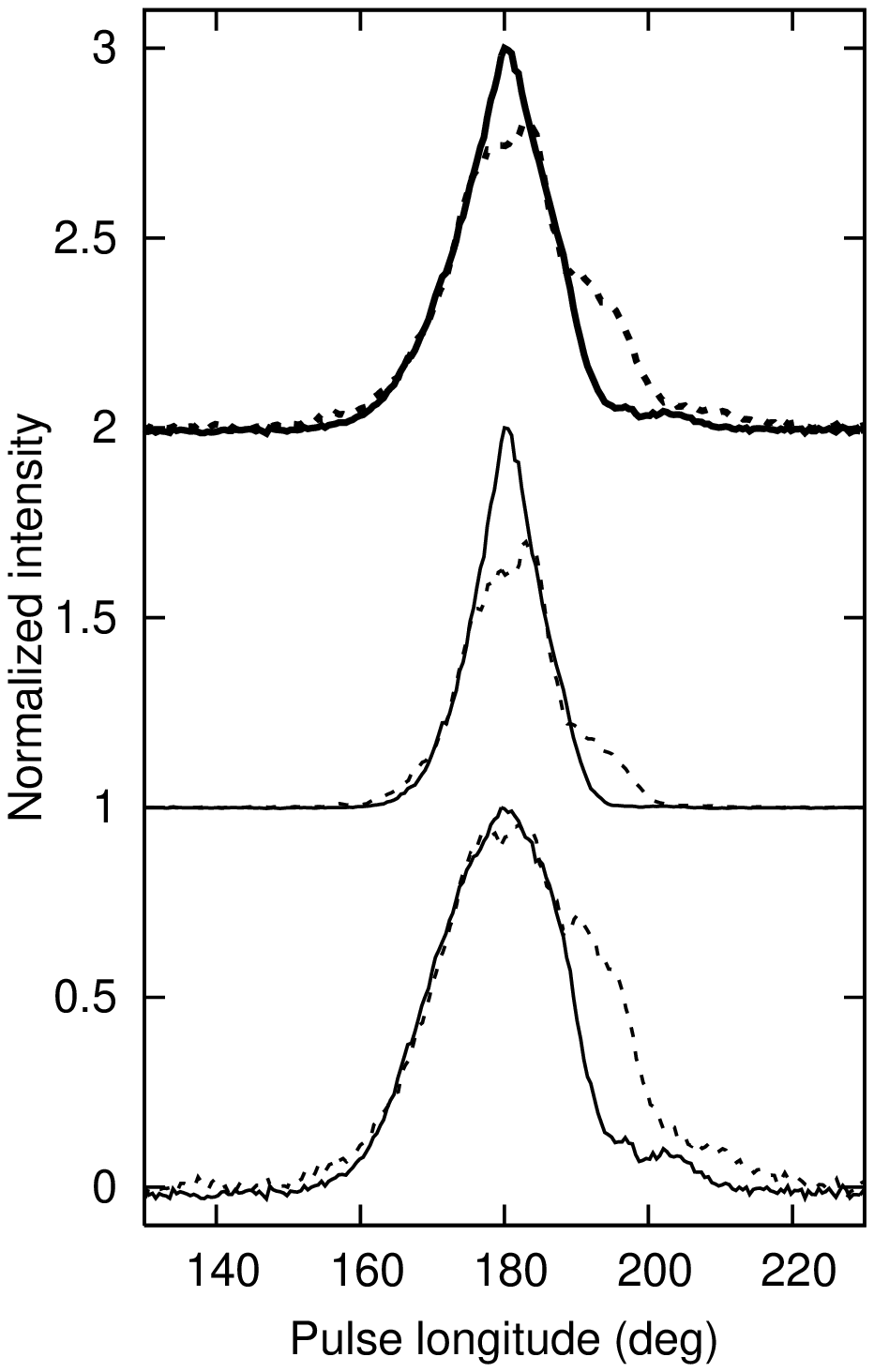}}
\end{center}
\caption{\label{separation_ao}{\bf Left:} The normalized total
emission pulse profile (thick solid line), which is decomposed into
the spiky (thin solid line) and weak emission profile (dotted line)
for the 327-MHz AO-P1 observation. {\bf Middle:} Same as the left
panel, but for the 1525-MHz AO-L observation. {\bf Right:} The
profiles of the total, spiky and weak emission with a vertical offset
of 2, 1 and 0 respectively of the AO-P1 (solid lines) and AO-L
observation (dotted lines). The peaks of the profiles of the AO-L
observation are normalized and those of the AO-P1 observation are
scaled such that the central peaks fit each other.}
\end{figure*}

\subsection{\label{SctSeparation2}Stability of the pulse profile revisited}

The reason why the pulse profile is unstable at low frequencies (see
Fig. \ref{Profiles}) is the presence of the spiky emission. These
spikes have a very uneven longitude distribution and therefore many
pulses are required to obtain a steady profile. This is demonstrated
in Fig. \ref{ProfilesSep}, where the profiles of successive blocks of
one thousand pulses each are plotted separately for the spiky and weak
emission. One can clearly see that it is the spiky emission that is
highly unstable, whereas the weak emission converges rapidly to a
stable profile. Moreover in this plot one can see that the observed
gradual change in the pulse profile in the AO-P2 (middle right panel
of Fig. \ref{Profiles}) is caused by a gradual change in the spiky
emission while the weak emission remains the same. Around pulse 11,000
(block 11) the profile of the spiky emission evolves from a peaked
profile to a broader profile that resembles more the shape of the
profiles of the weak emission. The fact that the profile of the spiky
emission can evolve independently of the weak emission is an
additional argument that PSR B0656+14 has two kinds of emission. This
also makes, as discussed in Sect. \ref{SctStability}, an instrumental
effect a very unlikely source for the observed profile shape changes.

\subsection{Profiles of the spiky and weak emission}

The pulse profiles of the separated spiky and weak emission are shown
in Fig. \ref{separation_ao}.  At both frequencies (left and middle
panels) there is a striking difference in shape between the spiky and
weak emission: the spiky emission profile is significantly narrower
more pointed than the total pulse profile, in contrast to the flatter
bell-like shape of the weak profiles. This is consistent with the
longitude-resolved energy distribution (Fig. \ref{lrced}) that shows
that the strongest spikes are in the central part of the pulse
profile.  At 327 MHz (left panel) the profile is generated by a mix of
both spiky and weak emission, but the ratio of their intensities is
different in the different profile components. At 1525 MHz (middle
panel) the profile of the spiky emission is an almost perfect
isosceles triangle, while the weak emission is much less symmetric and
only roughly triangular. At both frequencies the shoulder starting at
pulse longitude \degrees{200} is completely generated by weak
emission.

In the right panel of Fig. \ref{separation_ao} we have overlaid the
total, spiky and weak emission profiles of the AO-P1 and AO-L
observations. As in Fig. \ref{ProfilesOverlayed}, the centroid of the
central component is fixed at \degrees{180} and the profiles are
scaled in intensity such that the central component overlaps. Neither
the spiky nor the weak emission appears to evolve at the leading edge,
and the broadening of the profiles at low frequencies of both the
spiky and weak emission is only due to the appearance of more emission
components.

\begin{figure}[tb]
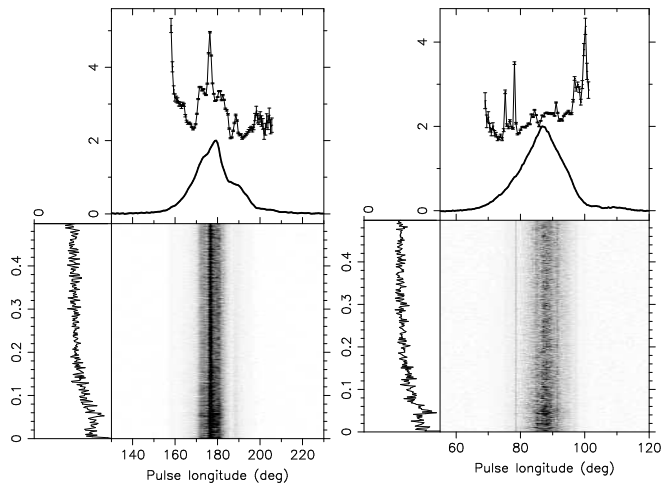

\resizebox{0.49\hsize}{!}{\includegraphics[angle=270]{5572f15a.ps}}
\resizebox{0.49\hsize}{!}{\includegraphics[angle=270]{5572f15b.ps}}
\caption{\label{ModulationProfiles} 
The modulation properties of PSR B0656+14 in the 327-MHz AO-P2 (left)
and 1525-MHz AO-L (right) observation. The top panel shows the
integrated profile (solid line) and the longitude-resolved modulation
index (solid line with error bars). Below this panel the LRFS is shown.
The power in the LRFS is horizontally integrated, producing the
side-panel.
}
\end{figure}

\subsection{The modulation-index profiles}

To characterize the modulation properties we have used the Discrete
Fourier Transform (DFT) to calculate the Longitude Resolved
Fluctuation spectrum (LRFS; \citealt{bac70b}), which is used to
calculate the longitude-resolved modulation index
(\citealt{es02,es03b}). The LRFS is obtained by taking DFTs along
lines of constant pulse longitude in the pulse stack. The resulting
spectra (bottom panels of Fig. \ref{ModulationProfiles}) have pulse
longitude on the horizontal axis and $P_1/P_3$ on the vertical
axis. Here $P_3$ is the subpulse pattern repetition rate and $P_1$ the
pulse period.

The longitude-resolved variance $\sigma_i^2$ is computed by vertical
integration of the LRFS from which the longitude-resolved modulation
index $m_i=\sigma_i/\mu_i$ is derived (where $\mu_i$ is the average
intensity at longitude bin $i$. For more details we refer to
\cite{wes06}.  As one might expect from the spiky nature of the
emission of this pulsar, its modulation index is unusually high,
especially at low frequencies.  The modulation index increases from
slightly below 2 at the leading edge of the profile to about 3 at the
trailing edge in the 1525-MHz AO-L observation (right panel of
Fig. \ref{ModulationProfiles}). The modulation-index profile of the
1380-MHz WSRT-L observation (not shown) is very similar, but slightly
lower probably due to a small amount of residual RFI in the AO
observations.  Notice also that the spectra of both AO observations,
despite the large number of pulses, are still ``contaminated'' by a
few very bright pulses causing vertical stripes in the LRFSs and
spikes in the modulation-index profiles. At 327 MHz the
modulation-index profile has evolved to a ``W-shape'' which varies
between 2 and 5 (left panel of Fig. \ref{ModulationProfiles}).

\begin{figure}[tb]
\begin{center}
\resizebox{0.49\hsize}{!}{\includegraphics[angle=0,trim=0 0 0 0,clip=true]{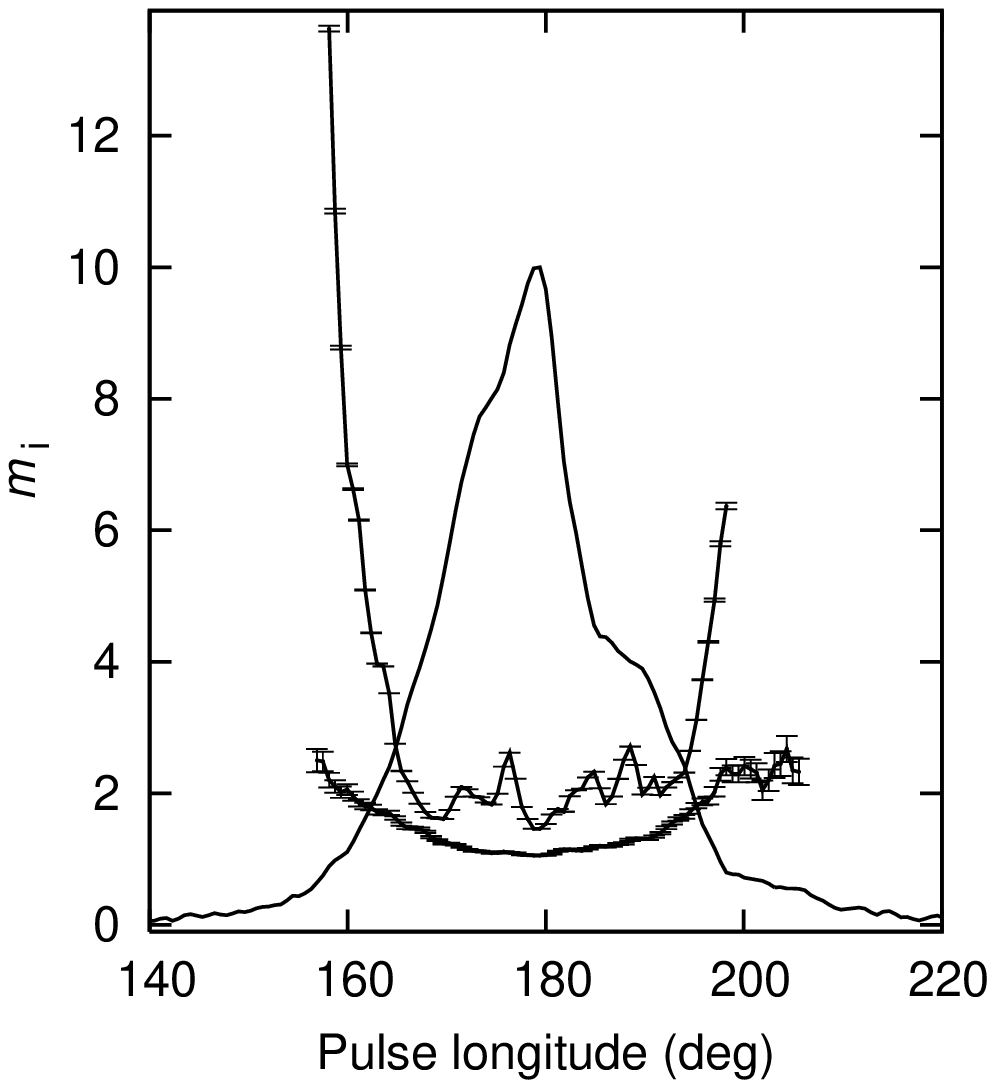}}
\resizebox{0.49\hsize}{!}{\includegraphics[angle=0,trim=0 0 0 0,clip=true]{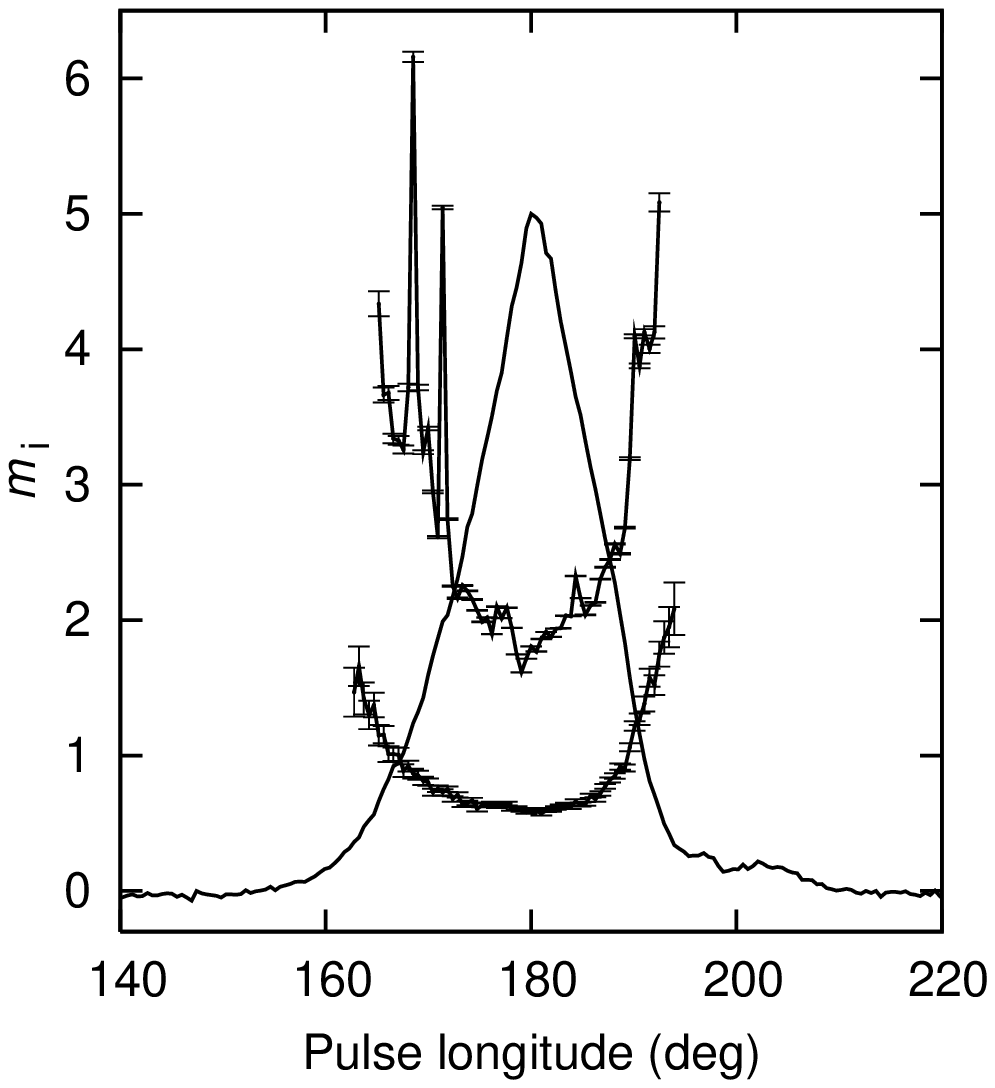}}
\end{center}
\caption{\label{ModulationSeparation}The left and right panels show
the 327-MHz AO-P2 observation and 1525-MHz AO-L observation
respectively. The modulation-index profiles of the spiky emission
(top) and weak emission (bottom) are superimposed on the pulse
profiles. The shapes of the modulation-index profiles are
important, while their magnitudes have little physical
meaning. Therefore the modulation-index profiles of the spiky emission
are scaled-down to make the profiles of comparable magnitude.} 
\end{figure}

In Fig. \ref{ModulationSeparation} the modulation-index profiles are
shown for the separated spiky and weak emission. The separation shows
that the modulation index of the spiky emission is higher and that of
the weak emission lower compared with the modulation index of the
total emission. The precise degree of this effect depends strongly on
the threshold intensity used for the separation. Therefore the shape
of the modulation-index profiles is more interesting than its exact
magnitude. To make the comparison in shape more clear in
Fig. \ref{ModulationSeparation}, we scaled-down the modulation-index
profile of the spiky emission.  It is striking that the modulation
profiles of both the spiky and weak emission at both frequencies have
a much simpler U-shape compared with the modulation index of all the
emission. A U-shape is usual for profile components (\citealt{wes06}),
suggesting that the profiles of the weak and spiky emission can be
regarded as true (overlapping) profile components. With this in mind
we can understand the complex shape of the modulation profiles of the
total emission in Fig. \ref{ModulationProfiles}.

Because the profile of the weak emission is broad, the edges of the
profile are dominated by the weak emission at 327 MHz (see left panel
of Fig. \ref{separation_ao}). This causes the outer edges of the
modulation-index profile to flare out at this frequency, like the
modulation-index profile of the weak emission (see
Fig. \ref{ModulationSeparation}). In the central part of the profile a
significant fraction of the total emission is spiky emission, so that
the local increase in the modulation index makes the modulation-index
profile of the total emission W-shaped.

At 1525 MHz the situation is slightly different because the centroid
of the profile of the weak emission has an offset toward the the
leading side of the pulse profile, while the centroid of the profile
of the spiky emission is exactly in the centre of the profile (see
middle panel of Fig. \ref{separation_ao}). Hence the weak emission
dominates the leading edge, but less so on the trailing edge.  This
fact, in combination with a relative low modulation index of the spiky
emission, causes the otherwise expected W-shape of the
modulation-index profile of the total emission to be blurred out by
the flaring of the modulation-index profile of the weak emission on
the trailing side of the profile.

\subsection{\label{SctBursting}\label{SctModulation}The quasi-periodic modulation}

In the LRFSs (bottom panels of Fig. \ref{ModulationProfiles}) one can
clearly see a $P_3=20\pm1P_1$ modulation feature, similar to that
observed by \cite{wes06}. The relatively large uncertainty in $P_3$
reflects the fact that this quasi-periodic modulation is weak compared
with the non-periodic intensity modulation (in the side panels of the
spectra in Fig. \ref{ModulationProfiles} there is only a relatively
weak and broad feature around $P_1/P_3=0.05$). Such a feature could be
caused by an intensity modulation or a phase modulation (drifting
subpulses). To distinguish between these possibilities, we calculated
the Two-Dimensional Fluctuation Spectrum (2DFS; \citealt{es02}). The
2DFS is obtained by taking DFTs along lines with different slopes in
the pulse stack. For a more comprehensive description of the methods
used we refer to \cite{wes06}. There is no evidence, in any of the
four observations, that there exists a preferred drift sense. A
preferred negative drift sense has been reported by \cite{bac81} at
430 MHz, but we suspect that was an artifact of a too short sequence
of pulses. We therefore conclude that the modulation features are the
result of an intensity modulation on a timescale of $P_3=20\pm1P_1$
with the phase shifts between the individual pulses essentially
random.

\begin{figure}[tb]
\begin{center}
\resizebox{0.49\hsize}{!}{\includegraphics[angle=0,trim=0 0 0 0,clip=true]{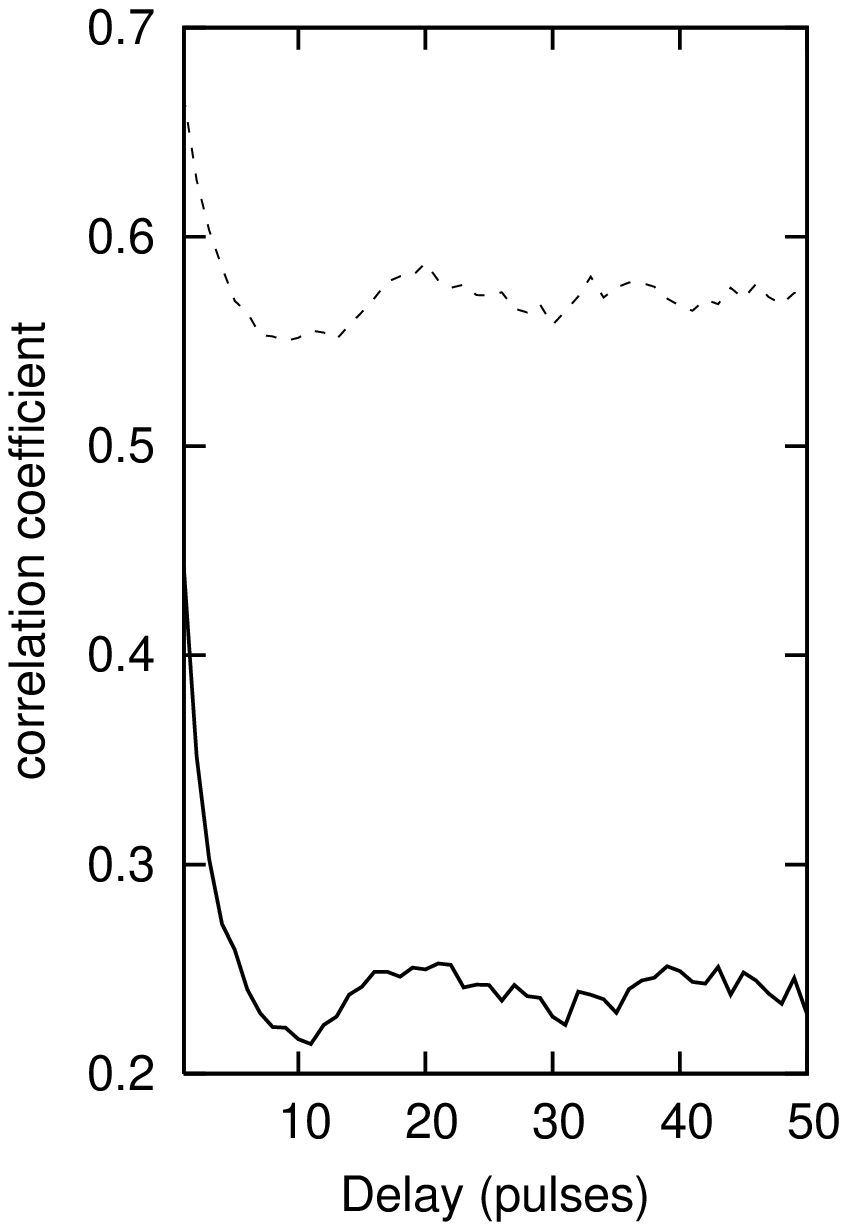}}
\resizebox{0.49\hsize}{!}{\includegraphics[angle=0,trim=0 0 0 0,clip=true]{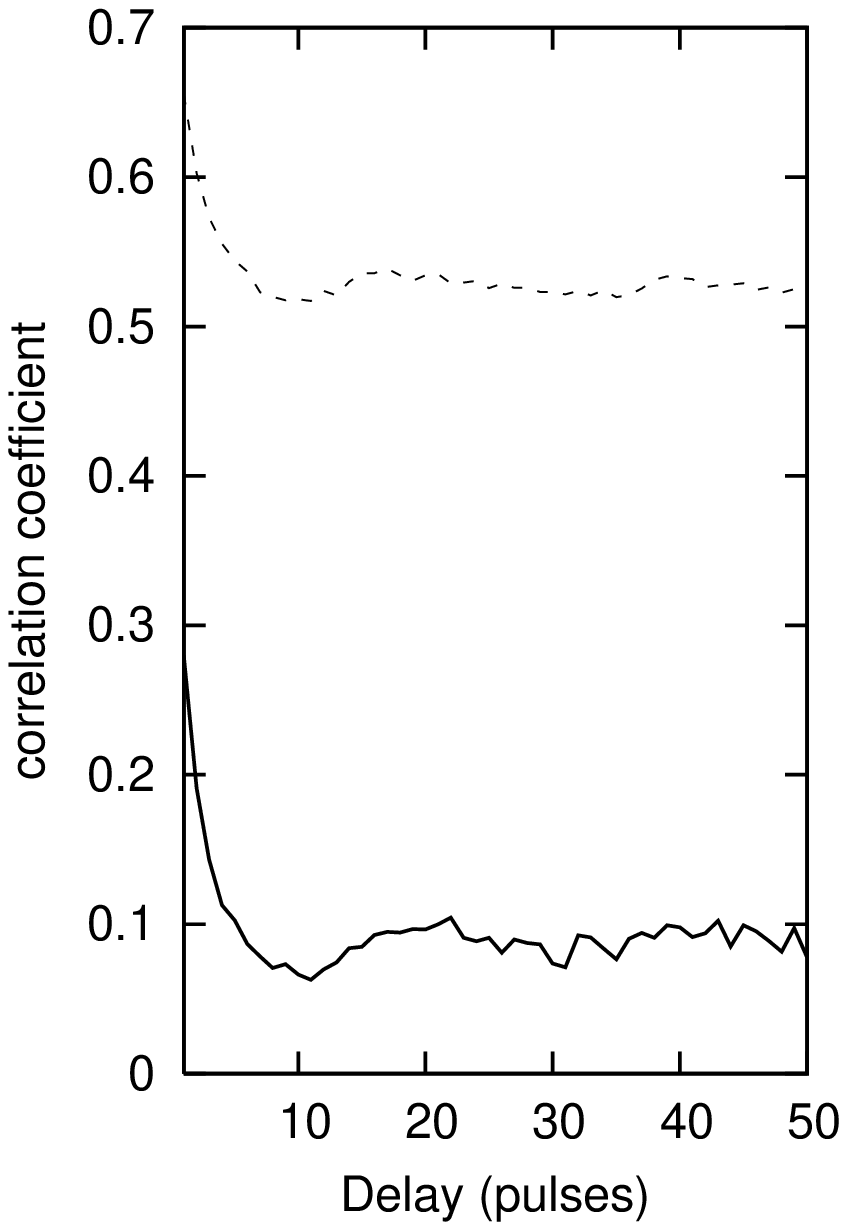}}
\end{center}
\caption{\label{Autocorr_pen}{\bf Left:} The ACF of the sequence of
pulse energies in the 327-MHz AO-P1 observation (solid line) and the
1380-MHz WSRT-L observation (dotted line). {\bf Right:} The ACF of the
sequence of pulse energies for the spiky emission (solid line) and the
weak emission (dotted line) in the 327-MHz AO-P1 observation. }
\end{figure}

To further explore this modulation feature, we calculated the
auto-correlation function (ACF) of the sequence of pulse energies
(left panel of Fig. \ref{Autocorr_pen}).  Both the 327-MHz AO-P1 and
the 1380-MHz AO-L observation show a dip in the ACF at a delay of
$\sim$10 $P_1$. Note that the ACFs of all observations show some
evidence for a slight increase at a delay of $\sim$20 $P_1$ and a
decrease at a delay of $\sim$30 $P_1$. This argues that there is a
weak quasi-periodicity with a period of $\sim$20 $P_1$ (which can just
be discerned in the pulse stack Fig. \ref{Stack}). Interestingly,
positive correlation with a delay of $\sim$20 $P_1$ is weak in
comparison with the dip at a delay of $\sim$10 $P_1$. This implies
that a burst of strong emission is not necessarily followed $\sim$20
$P_1$ later by another strong burst. Rather it shows that the burst
has decayed within $\sim$10 $P_1$.  Therefore the $P_3\sim20P_1$
feature that appears in the LRFS is only partially associated with a
periodic signal, but mostly with a characteristic timescale of the
rise and decay of the bursts (i.e. the bright pulses tend to cluster
as we discussed in Sect. \ref{SctSinglePulse}). The ACF of the WSRT-L
observation (as well as that of the not shown AO-L observation) is
flatter, which shows that at high frequencies there are less very
bright pulses and that the clustering of the bright pulses in a few
successive pulses is less.

In the right panel of Fig. \ref{Autocorr_pen} one can see that the dip
in the ACF at a delay of $\sim$10 $P_1$ is visible both in the spiky
and the weak emission. We attempted to test the hypothesis that only
the spiky emission, and not the weak emission, is responsible for the
dip. However, as the technique used could not wholly separate the
spikes from the weak emission, we could not tell whether the dip in
the ACF of the weak emission is caused by the spiky emission mixed in
the weak emission, or is an intrinsic property of the weak
emission. Nevertheless the dip in the ACF of the spiky emission proves
unambiguously that clusters of such emission are sustained over a
timescale of $\sim$10 $P_1$.

\begin{figure}[tb]
\begin{center}
\resizebox{0.7\hsize}{!}{\includegraphics[angle=0,trim=0 0 0 0,clip=true]{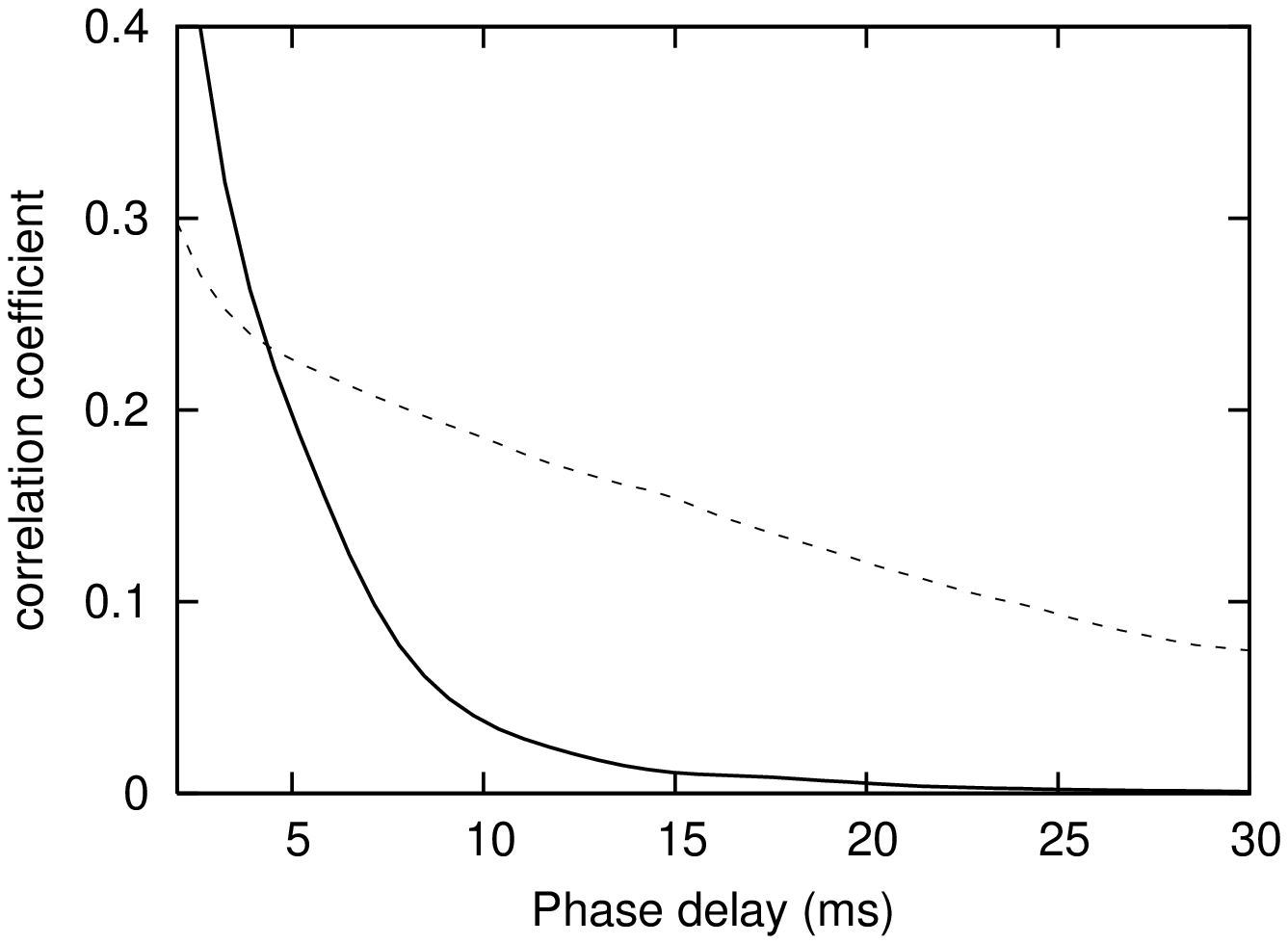}}
\resizebox{0.7\hsize}{!}{\includegraphics[angle=0,trim=0 0 0 0,clip=true]{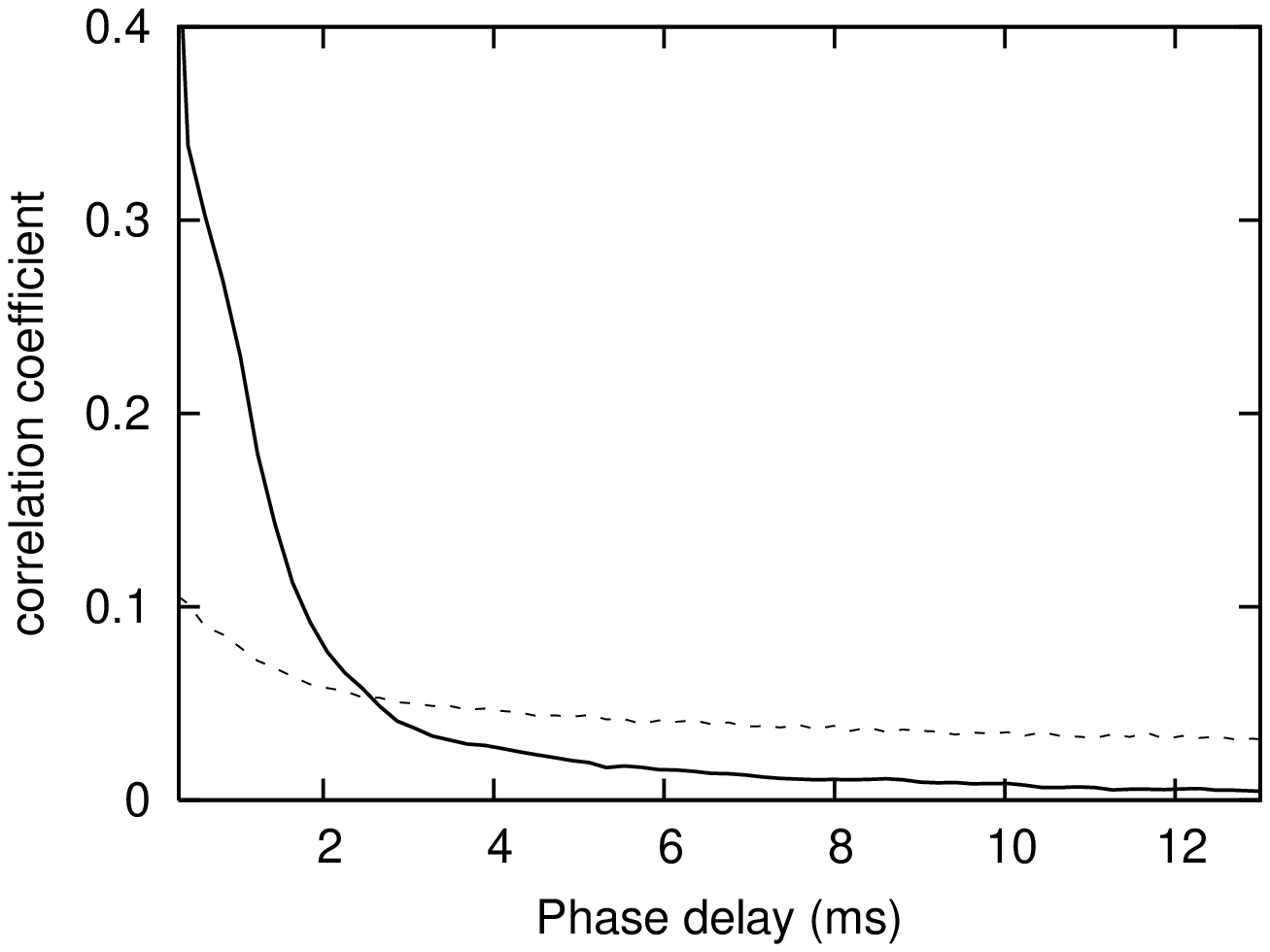}}
\end{center}
\caption{\label{ACFSeparation}The average ACF of the single pulses of
the spiky emission (solid line) and the weak emission (dotted line) in
the 327-MHz AO-P2 (top) and 1380-MHz WSRT-L observation (bottom). }
\end{figure}

\subsection{The time structure of the individual pulses}

In Fig. \ref{BrightestPulses} it is evident that the brightest pulses
have time-structures on two timescales. When multiple spikes appear in
a single pulse, they appear to be quasi-periodic with a timescale of
the order of 10 ms (this is clearest in the left panel of
Fig. \ref{BrightestPulses}). The spikes themselves show
micro-structure on a timescale of about 1 ms (only visible in the
right panel of Fig. \ref{BrightestPulses}). However, we need to know
if these timescales are typical for all emission and whether this
points to further differences between spiky and weak emission. We
therefore calculated the ACF of the single pulses and averaged them
over each of our four long observations.

In Fig. \ref{ACFSeparation} the resulting ACFs for the AO-P2 (top) and
WSRT-L observation (bottom) are shown. The 1-ms micro-structure is
clearly visible as a bump in the ACF of the spiky emission in the
WSRT-L observation. These are unresolved in the AO observations
because of the longer sampling time. At a delay of about 2 ms the ACF
of the WSRT-L observation flattens out, showing that the spikes become
unresolved on a timescale of a few ms (i.e. the typical width of the
spikes is a few ms). In the AO-P2 observation one can see no clear
bump in the ACF of the spiky emission at a delay around 10 ms, so this
value cannot be typical. In shorter stretches of data the ACF shows
various timescales for the separation between the spikes, which are
smeared out in the ACF of the whole observation in the range of about
5--10 ms.  Therefore there seems to be no fixed timescale for this
periodicity that is stable over a timescale of hours.

Both observations in Fig. \ref{ACFSeparation} show that at short
delays the weak emission correlates more weakly than the spiky
emission, but on longer timescales ($\gtrsim$ 3--5 ms) this is
reversed. The fact that the ACFs of the weak emission are very flat on
all timescales demonstrates again that the weak emission of a single
pulse is distributed over large parts of the profile width, yielding a
large cover fraction.  This is further support for the conclusion (as
suggested in Sect. \ref{SctSeparation} and \ref{SctSeparation2}) that
the PSR B0656+14's emission can be characterized as a combination of
weak emission with a broad longitude coverage across the profile and
thinly-distributed bright spikes.

\section{\label{SctDiscussion}Discussion and conclusions}

\subsection{Comparison with the known population of pulsars}

We have shown that PSR B0656+14 intermittently emits pulses that are
 extremely bright compared to normal pulsars and with pulse energies well above
10 {\Eav} these pulses formally qualify as giant pulses. Nevertheless
these pulses differ from giant pulses and giant micropulses in
important ways:

\begin{itemize}
\item The pulse-energy distribution of PSR B0656+14 does not show a
power-law tail (this is possibly also true for PSR B1133+16;
\citealt{kkg+03})
\item The widths of the bright pulses of PSR B0656+14 are much greater
than those of the extremely narrow giant pulses
(e.g. \citealt{spb+04}), although similar to that of the giant
micropulses.
\item The brightest pulses of PSR B0656+14 are an order of magnitude
brighter (both in integrated energy and in peak flux) than the giant
micropulses of the Vela pulsar (\citealt{jvkb01}) and PSR B1706--44
(\citealt{jr02}). 
\item The bright pulses of PSR B0656+14 are not strongly constrained
in pulse longitude. In most other pulsars, both fast
(e.g. B1937+21, B1821--24) and slow (e.g.  B0031--07) the pulse
longitudes where giant pulses appear are highly constrained, whether
towards one edge or centrally (e.g
\citealt{cst+96,kt00,rj01,ke04}). Also the giant micropulses detected
for the Vela pulsar and the young pulsar B1706--44 are highly confined
(in both cases to the leading edge of the pulse profile).
\item The claimed empirical correlation between the appearance of
giant pulses and high light-cylinder magnetic field strengths clearly
fails for PSR B0656+14 since its value (770 Gauss) is well below those
of most pulsars exhibiting giant pulses (around $10^5$ Gauss;
\citealt{cst+96}). However, PSR B0031--07 and a number of other slow
pulsars also easily fail this test, so the correlation may only be
valid for millisecond pulsars.
\item The bright pulses of B0656+14 do not coincide with any of the
observed X-ray peaks, in contrast to those of the well-known pulsars
exhibiting giant pulses. 
\end{itemize}

Despite the differences between the spiky emission of PSR B0656+14 and
the giant micropulses listed above, there are also interesting
parallels between PSR B0656+14 and Vela. The pulse profile
of PSR B0656+14 has a shoulder on the trailing edge reminiscent of a
similar feature discovered in Vela (\citealt{jvkb01}) and B1937+21
(\citealt{cst+96}). In both, the shoulder is not associated with the
giant pulses (\citealt{kt00}), and that also appears to be the case in
PSR B0656+14. Furthermore both pulsars have low-frequency intensity
modulations and neither pulsar shows a coincidence between the
locations of the radio and high-energy emission.

Many of the exceptional properties of PSR B0656+14 have led us to
point out (\citealt{wsr+06}) that this pulsar, were it not so near,
could have been discovered as an RRAT. Chief among these are the
infrequent, but luminous pulses with very high peak fluxes and widths
comparable with those of the RRATs. Furthermore, like RRATs, PSR
B0656+14 has a light-cylinder magnetic field strength which is not
exceptional.

Surprisingly, the middle-aged PSR B0656+14 also shares many properties
with recycled millisecond pulsars. For instance both have wide
profiles that have a more or less constant width or component
separation over a very wide frequency range
(e.g. \citealt{kll+99})\footnote{However, it should be noted that
invariant profile widths is not a property confined to millisecond
pulsars. \citep{mr02a} have shown that a whole class of bright slow
pulsars have component widths and component separations independent of
frequency.}. Furthermore the emission of millisecond pulsars is
usually highly polarized with a flat linear polarization profile
(e.g. \citealt{xkj+98}). PSR B0656+14 is 92\% linear polarized at 400
MHz and also the position angle is remarkably flat (\citealt{gl98}).
An interesting comparison might be drawn between the B0656+14 and a
number of millisecond pulsars which are also known to have
slowly-varying unstable profiles (e.g. \citealt{rk03,kxc+99,bs97}),
possibly resulting from spiky emission in one or more components.

\subsection{The spiky emission}

We show in this paper that the emission of PSR B0656+14 can be
characterized by spiky emission that is not confined to a narrow
pulse-longitude range, but generally concentrated toward the centre of
its profile.  This spiky emission has a low occurrence rate
within each pulse and is often quasi-periodic in structure. In
addition, pulses with spiky emission have a tendency to cluster in
successive pulses. By contrast, there is also a weak emission, which
appears with a high occurrence rate over the full width of the pulse
and varies little from pulse to pulse. Furthermore, the profiles of
the spiky and weak emission have very different shapes and frequency
evolution. Taken together, these properties all argue for two distinct
types of emission.

Separate properties of weak and strong emission can be supported by
the fits of the pulse-energy distribution. However, in
Fig. \ref{lred_fit} one can see that all pulses with a peak-flux
weaker than the average are below the noise level. Therefore more
sensitive observations (with at least an order of magnitude increase
in $S/N$) are required to see single weak pulses and to find out if
there exists a break in the pulse-energy distribution.  Such
an observation would show if the spiky and weak emission are
associated with the extreme ends of a single smooth energy
distribution, or if they are two physically distinct distributions.

Pulse profiles of many pulsars can accurately be decomposed into a
small number of Gaussian emission components (\citealt{kwj+94}).  This
is not possible for the isosceles shape of the profile of the spiky
emission at high frequencies. This, together with the width of the
central component that is invariant with frequency, may suggest that
it reflects a permanent geometrical feature. It might be related to an
intensity-enhancing (e.g. caustic) effect that boosts the otherwise
weak emission.  Due to the low occurence rate of the spiky emission,
the profile requires an unusually long timescale to achieve stability
(over 25,000 pulses at 327 MHz). Despite the high variability in shape
of the spiky emission profile on short timescales, we found that the
radio power output remains remarkably stable and this may indicate
that processes ``repackage'' the power within the magnetosphere,
rather than generate it.

The emission of PSR B0656+14 shows a quasi-periodicity of $20P_1$,
which is intensity modulated, rather than phase modulated. Similar
fluctuations without drift are often associated with core emission
within a conal structure (e.g. PSR B1237+25; \citealt{sr05}). To the
best of our knowledge, there have been no theoretical attempts to
account for the periodicity of this emission. Possibly the modulation
should be considered in the framework of the charge flow balance
within the entire magnetosphere. Although we conventionally suppose in
such a pulsar that we only have a view of one pole, it has recently
been plausibly suggested (\citealt{dfs+05}) that in selected stars
(all of them pulsed X-ray sources, as is PSR B0656+14) it is possible
to clearly detect radiation from particles downflowing towards the
pole that is hidden from our view.  The widely-separated high-energy
peaks located well away from the phase of the polar cap point to
activity in the outer magnetosphere, possibly associated with an outer
gap. Also in the case of PSR B0656+14 we may be dealing with emission
which is stimulated, either directly on indirectly, by an inflow - as
well as an outflow - of charged particles (an analogy with rainfall is
not inappropriate), with short-lived showers of particles
intermittently injected into the polar regions in a quasi-periodic
manner. This idea will be explored in a subsequent paper, in which we
will also present polarization data and attempt to link the radio
emission to the high-energy peaks.

\begin{acknowledgements}
GAEW and JMR thank the Netherlands Foundation for Scientific Research
(NWO) and the Anton Pannekoek Institute, Amsterdam, for their kind
hospitality, and GAEW the University of Sussex for a Visiting
Fellowship. Portions of this work (for JMR) were carried out with
support from US National Science Foundation Grants AST 99-87654 and
00-98685. Arecibo Observatory is operated by Cornell University under
contract to the US National Science Foundation. The Westerbork
Synthesis Radio Telescope is operated by the ASTRON (Netherlands
Foundation for Research in Astronomy) with support from NWO.

\end{acknowledgements}

\end{document}